\documentclass[hidelinks,11pt,a4paper]{article}
\usepackage{caption,jheppub}
\usepackage{graphicx}
\usepackage{sidecap}
\usepackage{dcolumn}
\usepackage{bm}
\usepackage{amsmath,amsfonts,amssymb}
\usepackage{slashed}
\usepackage{braket,xcolor}
\usepackage{verbatim}
\usepackage{tikz}
\usepackage{comment}

\usepackage{geometry}
\usepackage{physics}
\usepackage{xparse}
\usepackage{amsmath}
\NewDocumentCommand{\tens}{t_}
 {%
  \IfBooleanTF{#1}
   {\tensop}
   {\otimes}%
 }
\NewDocumentCommand{\tensop}{m}
 {%
  \mathbin{\mathop{\otimes}\displaylimits_{#1}}%
 }
\usepackage{array}

\usepackage{caption}
\usepackage{subcaption}
\usepackage{amsthm,amsfonts,dsfont,tensor,stmaryrd,manfnt,mathrsfs,svrsymbols,mathtools}
\usepackage{subcaption}
\usepackage{amssymb}
\usepackage{multirow}
\usepackage{float}
\usepackage{graphicx}
\usepackage{amsmath}
\usepackage{placeins}
\usepackage{amsfonts,mathtools,resizegather}
\usepackage[utf8]{inputenc}
\usepackage{csquotes}
\usepackage{comment}
\usepackage{booktabs}
\usepackage{enumitem}
\renewcommand{\chi}{\zeta}

\newcommand{\cC}{\mathcal{C}}
\newcommand{\cD}{\mathcal{D}}
\newcommand{\cK}{\mathcal{K}}
\newcommand{\cL}{\mathcal{L}}
\newcommand{\cN}{\mathcal{N}}
\newcommand{\cH}{\mathcal{H}}

\newcommand{\Z}{\mathbb{Z}}

\newcommand{\R}{\mathbb{R}}
\newcommand{\C}{\mathbb{C}}
\newcommand{\SL}{\mathrm{SL}}

\newcommand{\bchi}{\bar{\chi}}

\newcommand{\bxi}{\bar{\xi}}

\newcommand{\bzeta}{\bar{\zeta}}

\newcommand{\osp}[1]{\mathfrak{osp}(#1)}
\newcommand{\su}[1]{\mathfrak{su}(#1)}
\renewcommand{\sl}[1]{\mathfrak{sl}(#1)}

\usepackage{soul}
\newcommand\quotes[1]  {`{#1}'}
\newcommand{\gev}[1]{\left\langle#1\right\rangle_{\textrm{sca}}}

\title{Spread complexity for the planar limit of holography}

\author[a]{Rathindra N. Das,}
\author[b]{Saskia Demulder,}
\author[a]{Johanna Erdmenger,}
\author[c]{Christian Northe}

\affiliation[a]{Institute for Theoretical Physics and Astrophysics and\\ Würzburg-Dresden Cluster of Excellence ct.qmat,\\ Julius-Maximilians-Universität Würzburg, Am Hubland, 97074 Würzburg, Germany}
\affiliation[b]{Department of Physics, Ben-Gurion University of the Negev,\\ David Ben Gurion Boulevard 1, Beer Sheva 84105, Israel}
\affiliation[c]{CEICO, Institute of Physics of the Czech Academy of Sciences,\\
Na Slovance 2, 182 00 Prague 8, Czech Republic}

\newcommand\id		{\mathbf{1}}

\abstract{Complexity is a fundamental characteristic of states within a quantum system. Its use is however mostly limited to bosonic systems, inhibiting its present applicability to supersymmetric theories. This is also relevant to its application to the AdS/CFT correspondence. To address this limitation, we extend the framework of spread complexity beyond bosonic systems to include fermionic and supercoherent states. This offers a gateway to compute spread complexity analytically for any semiclassical system governed by a Hamiltonian associated with a Lie (super)algebra.
This requires extending the Krylov chain to a \textit{Krylov path} in a higher-dimensional lattice. A detailed analysis of supercoherent states within the super Heisenberg-Weyl and OSp$(2|1)$ algebras elucidates distinct contributions from bosonic and fermionic degrees of freedom to the complexity. This generalisation allows us to access the semiclassical regime of the planar limit of the holographic correspondence. We then compute the spread complexity of large charge superstring states on the gravity side, which are equivalent to the dual gauge states. The resulting complexity leads to Krylov paths capturing the geometry in which the string propagates.
}

\begin{document}
\maketitle
\flushbottom

\newpage

\section{Introduction}

The notion of complexity has gained growing attention in recent years within the high-energy theory community. Initially proposed as capturing the growth of the Einstein-Rosen bridge for AdS black holes \cite{Susskind:2014rva}, complexity has grown to become a promising tool to probe a system's dynamics. In particular Krylov complexity, and more recently spread complexity, were used to investigate quantum chaos in \cite{Parker:2018yvk} and \cite{Balasubramanian:2022tpr}, respectively. The scope of applicability of complexity in the Krylov space extends from quantum field theories~\cite{Avdoshkin:2022xuw,Caputa:2023vyr,He:2024xjp,He:2024hkw,Camargo:2022rnt} to the study of quantum phase transitions~\cite{Caputa:2022eye,Caputa:2022yju,Bhattacharya:2024hto}.

A logical and necessary development is to understand the role of complexity in superstring theory and most notably in the AdS/CFT correspondence which requires to take into account both bosonic and fermionic contributions. This is what we set out to investigate in the present article. Spread complexity, which was first defined in \cite{Balasubramanian:2022tpr}, measures the growth of a state under Schr\"odinger time evolution.  Often, the numerical challenge posed by the Lanczos algorithm, which is a key component in obtaining spread complexity, is a major hurdle towards extracting the spread complexity of any chosen quantum system.  However, in \cite{Caputa:2021sib} it was already observed that for a particular class of states, namely coherent states, spread complexity can be computed purely analytically, removing in this case the need to apply the Lanczos algorithm. By identifying the Hamiltonian with the operator driving the coherent state, the authors of \cite{Caputa:2021sib} identified the Krylov basis with states in a highest weight representation.

Coherent states, being the most ``classical'' quantum states, are prevalent when studying the semiclassical properties of quantum systems. 
In string theory and holography, coherent states are known to capture e.g.~D-brane physics, Lin-Lunin-Maldacena (LLM) geometries \cite{Lin:2004nb,Berenstein:2004kk}, and the motion of strings with large quantum charges -- the latter is a central topic in the second part of this article. Crucially, in its current formulation, spread complexity is too limited to address most of these holographic scenarios. More specifically, the Hilbert space of states driven by a Hamiltonian is well-approximated in the semi-classical limit by a basis of coherent states. 

In this paper, we construct a direct generalisation of spread complexity, taking the initial state as the coherent state itself. This state is driven by the relevant semiclassical expectation value of the Hamiltonian, which generically is not linear in the algebra generators. 
Extracting the spread complexity of a given operator requires significantly extending the present formulation of spread complexity. Indeed, although spread complexity is by now thoroughly understood for coherent states with rank-one Lie algebras, in general holography necessitates the addition of fermionic generators and understanding coherent states associated to semisimple Lie groups. We study precisely this extension. Our main aim is to obtain spread complexity for superstring states and their duals in the semiclassical section of the planar limit of holography.

The AdS/CFT correspondence offers a unique opportunity to probe gauge and gravity theory simultaneously. In particular in the so-called planar limit \cite{tHooft:1973alw}, by taking $N\rightarrow \infty$, the correspondence becomes especially amenable since both sides are integrable. In the planar limit, the relevant processes are dominated by planar Feynman diagrams that correspond to single-trace operators in the gauge theory. In \cite{Minahan:1990sc}, the authors made the crucial observation that in this limit, when restricting to scalar operators, the one-loop planar dilatation operator can be identified with an integrable spin chain. The task of determining the anomalous dimensions is then tremendously simplified by using integrability techniques, in particular applying the Bethe ansatz. In many cases however, one has to rely additionally on supersymmetry, i.e. BPS states or the corresponding Berenstein-Maldacena-Nastase (BMN) limit \cite{Berenstein:2002jq}, in order to make explicit computations and probe the ramifications of the holographic correspondence.

In \cite{Frolov:2003qc} it was realised that a further simplification occurs when taking an additional semiclassical limit offering a unique match between the strings propagating in $AdS_5\times S^5$ with the anomalous dimensions of the corresponding Super Yang-Mills (SYM) operators. These expectations are strikingly met in \cite{Kruczenski:2003gt,Kruczenski:2004kw,Kruczenski:2004cn}. Taking the semiclassical limit establishes a direct map between near-BPS states on the two sides of the duality. Specifically, this regime relates the phase space action of point-like string states carrying large quantum numbers (``fast string'') to the coherent state representation of the spin chain describing SYM.  This correspondence does not only relate states on the gauge with states on the gravity side, but identifies the non-linear sigma-model? action capturing the dynamics on both sides. In other words, it relates not only between the dimension of the SYM operator and energy of string states but establishes a direct relation between the collection of coherent states on the SYM side and string profiles on the gravity side.  By this direct matching of Lagrangians on both sides, the energies of a large class of string solutions match with the corresponding anomalous dimensions of gauge theory operators, without the need to compute these case-by-case. By now this duality relation has been thoroughly tested, see the reviews \cite{Tseytlin:2004cj,Tseytlin:2004xa,Minahan:2010js} for an exhaustive bibliography on the subject.

A key challenge is to identify a measure of complexity that is compatible with holography. Indeed, although multiple definitions of complexity have been put forward in recent years, both on the field theory and gravity sides, whether and how the different definitions match across the correspondence is an active area of research~\cite{Brown:2015lvg,Carmi:2017jqz,Susskind:2014rva,Susskind:2018pmk,Yang:2019alh,Engelhardt:2021mju,Couch:2016exn,Erdmenger:2021wzc,Erdmenger:2022lov,Brown:2015bva,Belin:2022xmt,Belin:2021bga,Caputa:2024sux,Rabinovici:2023yex,Fan:2024iop,He:2024pox,Das:2024zuu,Bhattacharya:2024szw,Balasubramanian:2024lqk}.  Here, by specialising to the particular sector of the holographic correspondence described above, we simultaneously   compute spread complexity for strings propagating in $AdS_5\times S^5$ as well as their dual states, described as ferromagnetic excitation of the spin chain. 
It is in this regime of the holographic correspondence that we extend the framework of spread complexity and apply it to string states and their duals in the semiclassical limit of planar holography. For the first time, this allows us to access spread complexity of string states.

To set the stage, we begin by considering purely fermionic coherent states, demonstrating how statistics drastically affects the usual picture one is used to in spread complexity. Turning first to the fermionic Heisenberg-Weyl coherent state, leads, unsurprisingly to a finite rather than semi-infinite chain. This innocent fact however turns out to be one of the building blocks for the interpretation of spread complexity in more complicated setups. After considering the simple case of a single fermion, we discuss multi-fermion coherent states, a relevant ingredient when interpreting purely fermionic string states.

The situation rapidly becomes more involved, once setups containing fermions \textit{and} bosons are considered. 
In this situation the Krylov chain picture developed in \cite{Parker:2018yvk} needs to be refined. Indeed, we argue that the Krylov chain should be thought of as carving a Krylov path through a higher-dimensional lattice with bosonic and fermionic directions. While its bosonic directions are semi-infinite, the fermionic directions are finite. Furthermore, the Krylov path is generated by an emergent dynamical rank one algebra giving the Krylov path its \quotes{direction}. Such emergent algebras of rank one governing Krylov evolution has been observed already in \cite{Caputa:2024sux,Fan:2024iop}, yet not explained. Our studies in this paper allow us to pinpoint precisely by what mechanism these dynamical spread algebras arise. Concretely, we study two examples: Firstly, a coherent state generated from one free boson and one free fermion. Secondly, the more complicated setup of a supercoherent state governed by the supergroup $OSp(2|1)$. Choosing this supergroup is natural, because its bosonic subgroup $\sl{2}$ has been studied thoroughly in the literature, see for instance \cite{Caputa:2021sib}.

The example of the supergroup $OSp(2|1)$ also gives us an opportunity of presenting a spread driven by a fermionic operator, which does not give rise to finite-dimensional Krylov space. This runs counter to the expectations one might naively gather from considering free fermion modes only. We show that an evolution driven by a supercharge leads to an infinite-dimensional Krylov space. The reason for this is as simple as it is profound, namely that supercharges square into translations.

With these tools and insights at hand, we turn to compute spread complexity in the planar sector of the holographic correspondence, applying the semiclassical limit. As described earlier, in this limit, both sectors share the same dynamics, and a single computation yields a complexity that applies to both the string and gauge sectors simultaneously.

The crucial observation is that spread complexity is best approached from the gauge side, where the Hilbert space of states admits a coherent state description. One essential difference compared to spread complexity in traditional coherent states is that the initial state is no longer a single lowest-weight state but a \textit{product state of many coherent states}. Remarkably, string states propagating on different submanifolds of  $AdS_5 \times S^5$  lead to Lanczos coefficients corresponding to effective  $SL(2)$  or  $SU(2)$  coherent states. While surprising at first sight, this result aligns with the expectations built in the previous sections: a Krylov path through a higher-dimensional space emerges.

The key feature now is that, since the Hamiltonian is a spin chain Hamiltonian. This has two implications. Firstly, the evolution through the phase space of the coherent states follows now a different one-dimensional trajectory, fixing a different time-dependence than before. Secondly, taking the continuum limit the spin chain acquires a spatial coordinate. That the spread complexity depends now on space and time reflects the worldsheet dependence of the string states. Taking the string picture at face value, yet another pattern emerges: the spread complexity is bounded because the string's motion is restricted to a compact submanifold, even when embedded in the non-compact space of  $AdS_5 \times S^5$ .

In section \ref{sec:review_spread_C} we briefly review the salient feature of spread complexity together with two simple of crucial examples for what is to follow. In section \ref{sec:generalise_speadC}, we lay out how spread complexity generalises to coherent states associated to Lie group of higher rank and their super-extensions. We close this section by reviewing the semiclassical approximation by mean of the coherent state path integral, and explain its role in the current extension of spread complexity. In section \ref{sec:fermion_osc}, we analytically compute the spread complexity for coherent states generated from fermionic generators. Here we need to distinguish two cases: one where the coefficients are pure c-numbers or when they are allowed to be Grassmann-valued. We then show how to generalise this analysis to  multiple fermions. In section \ref{sec:superC}, we consider cases that include both bosonic and fermionic generators: supercoherent states. The key example is the $OSp(2|1)$-algebra, which we discuss in detail. Finally in section \ref{sec:semicl_spin_chains}, we apply the machinery to holographic spin chains and compute the spread complexity for rotating strings in different subsectors of the symmetry algebra.

\section{Review of spread complexity}\label{sec:review_spread_C}

\begin{figure}[t]
    \centering
    \includegraphics[width=0.52\linewidth]{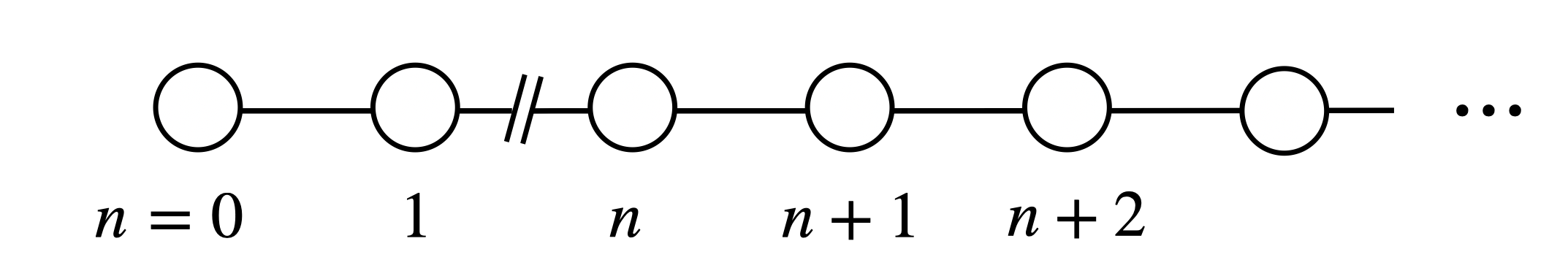}
    \caption{The Lanczos algorithm constructs a Krylov basis which leads to an auxiliary semi-infinite chain reflecting the gradual spread of the initial state over the Hilbert space.}
    \label{fig:simple_semi_chain}
\end{figure}

The notion of spread complexity was introduced in \cite{Balasubramanian:2022tpr} and quantifies the spread of a quantum state through Hilbert space upon evolution with an operator of choice, which we now review.  Consider a quantum system with Hilbert space $\cH$ governed by the Hamiltonian $H$ and consider state evolution in the Schr\"odinger picture, i.e. $|\psi(t)\rangle = e^{-iHt}|\psi(0)\rangle$. The evolved state is a linear combination of
\begin{equation}
|\psi\rangle,~ H|\psi\rangle,~ H^2|\psi\rangle,~ \dots,
\label{eq:basis state}
\end{equation}
where $|\psi(0)\rangle$ is denoted as $|\psi\rangle$. The $\cD$-dimensional subspace $\mathcal K\subseteq\cH$ spanned by the terms in \eqref{eq:basis state} is coined the Krylov space. Orthogonalising these states using the Lanczos algorithm, the resulting basis is called the Krylov basis and was shown in \cite{Balasubramanian:2022tpr} to select the complexity with minimal cost. We denote this basis by $|K_n\rangle$. The algorithm also produces a set of coefficients $(a_m,b_n)$, which, as we see shortly, characterise the time evolution of the initial state through Hilbert space, as measured by the Krylov basis. Expanding the Schr\"odinger state $|\psi(t)\rangle$ in terms of the Krylov basis yields
\begin{equation}
|\psi(t)\rangle = \sum_{n=0}^{\mathcal{D}-1}\psi_n(t)|K_n\rangle\,,
\label{eq:state in the Krylov basis}
\end{equation}
and substituting \eqref{eq:state in the Krylov basis} into the Schr\"odinger equation, we obtain\footnote{The closely related Krylov complexity, which describes operator, rather than state, growth, follows precisely the same procedure. The Heisenberg, rather than Schr\"odinger, evolution implies however that the equations will differ by a number of crucial signs, leading to a complexity which is quantitatively and qualitatively different.}
\begin{equation}
i\partial_t{\psi}_n(t) = a_n\psi_n(t)+b_{n+1}\psi_{n+1}(t)+b_n\psi_{n-1}(t)\,.
\label{eq:Krylov chain for state}
\end{equation}
The initial condition is $\psi_n(0)=\delta_{n0}$ by definition. Crucially for what is to come, we see that the weights $\psi_n$ of each Krylov basis element $|K_n\rangle$ can be interpreted as a wave-function with support on a semi-infinite chain, see fig. \ref{fig:simple_semi_chain}. How far, on average, the state $\ket{\psi}$ reaches into the chain -- or equivalently how many Krylov basis elements it has spread over -- is defined as its spread complexity, 
\begin{equation}
\mathcal C(t) \equiv \sum_{n=0}^{\mathcal{D}-1}n|\psi_n(t)|^2
\,.
\label{eq:Krylov state complexity}
\end{equation}
The complexity can be viewed as expectation value of a spread operator $K=\sum_nn\ket{K_n}\bra{K_n}$, i.e. $\mathcal  C(t)=\bra{\psi(t)}K\ket{\psi(t)}$. 
Note that any monotonous function $f(n)$ provides a good weight for $|\psi_n(t)|^2$ in this sum. It is customary to employ the simplest choice $f(n)=n$.

Unfortunately, the Lanczos algorithm is infamously unstable, prone to a quick build-up of numerical errors. In several cases this problem can be circumvented altogether. Most notably this occurs when the states under consideration are coherent states where the Hamiltonian is identified with the Hamiltonian driving the semiclassical evolution in phase space. This is discussed in depth in section \ref{sec:semiclassics}. Coherent states driven by operators lying in linear algebras of rank 1 were in fact already studied in \cite{Caputa:2021sib}. For such systems, the Lanczos coefficients and spread complexity can be computed analytically. These examples however remained somewhat artificial, as they do not directly correspond to physical systems\footnote{Note that in \cite{Caputa:2022eye}, the authors also considered a spin chain, but a clever transformation reduced the semiclassical Hamiltonian to a linear combination of algebra generators. In general such a transformation is not available.}. Indeed, taking the semiclassical approximation reviewed in subsection \ref{sec:semicl_spin_chains}, the resulting system rarely displays a Hamiltonian which is a simple linear combination of algebra generators. In fact, a well-known example, and the one studied here, are spin chains leading the semiclassical dynamics governed by a Hamiltonian valued in the tensor product of the algebra, generically of rank higher than one.

\section{(Super-)coherent states and semiclassics}\label{sec:generalise_speadC}
The string and gauge side of the $AdS_5\times$CFT$_4$-holographic correspondence is characterised by the $PSU(2,2|4)$ supergroup. In order to compute the spread complexity of string states the planar sector of holography first necessitates understanding the Hilbert space of states spanned by coherent states associated with this Lie group and its super subgroup.  Indeed, to generalise spread complexity for these semiclassical string states, we are faced with
\begin{itemize}
    \item[\textit{i.}] displacement operators taking values in a Lie group of rank larger than 1,
    \item[\textit{ii.}] displacement operators taking values in a \textit{super} Lie group.
\end{itemize}
In this section, we review, in turn, the construction of the associated coherent states and their Hilbert spaces. Along the way, we explain the critical ingredients to compute spread complexity for these states.

\subsection{Lie group-valued coherent states}
We first review the construction of coherent states for Lie group, as introduced by Perelomov and Gilmore \cite{perelomov1972coherent,Gilmore1972,Perelomov1986}.
Consider  a Lie group $G$, together with a unitary, irreducible representation $\rho$ acting on states $\ket{\psi}\in\cH$. For a chosen state $\ket{\psi_0}$ the set of coherent states is determined by the data $\{\rho,\ket{\psi_0}\}$ with vectors $\ket{\psi}\in \cH$ such that $\ket{\psi}=\rho(g)\ket{\psi_0}$ for $g\in G$. By restricting to connected and simply connected Lie groups, coherent states take on the form
\begin{align}
|\psi(\xi)\rangle = \exp\left(i \sum_\mu\xi_\mu T_\mu  \right)\ket{\psi_0}\,,
\end{align}
where  $\{T_\mu\}$ are generators of the Lie algebra $\mathfrak{g}$ of $G$ and $\xi_\mu$ are a priori complex parameters. Observe that $D(\xi)=e^{i \sum_\mu\xi_\mu T_\mu}$, referred to as the displacement operator or displacer in the following,  is unitary once one specialises to real parameters $\xi_\mu$. Denote by $G_{\psi_0}$ the stabiliser subgroup, which consists of elements $g\in G_{\psi_0}\subset G$ such that $\rho(g)\ket{\psi_0}=e^{i\phi}\ket{\psi_0}$ remains invariant up to a phase $\phi\in [0,2\pi)$. In particular, the set of coherent states is given by the elements of the coset space $G/G_{\psi_0}$.

A particular example of such coherent states is realised when the exponent of the displacement operators is chosen to be the Hamiltonian of a system, automatically identifying $\xi$ with the time parameter $t$. It is precisely this example that has been the object of study in e.g. \cite{Caputa:2021sib,Balasubramanian:2022tpr,Haque:2022ncl,Chattopadhyay:2023fob}, in the context of spread complexity. For many purposes, however, the displacement operator is not generated by the Hamiltonian of the system. In addition, the Lie algebra generator $\{T_\mu\}$ has so far only been considered to be valued in \textit{bosonic} Lie algebras of \textit{rank one}. As we will see, generalising beyond this simple setting requires a revision of the current formalism of Krylov complexity. 

Assuming the Lie algebra is semisimple, the algebra generators can always be organised into ladder operators $E_\alpha$ and commuting elements $H_i$\footnote{In representation theory, ladder operators are called roots $E_\alpha$ and the commuting elements are known as Cartan elements $H_i$ of the algebra. We do not use this nomenclature in this paper.}. 
Fixing a Lie algebra $\mathfrak{g}$ and a highest weight representation $\rho_\Lambda$, we declare its corresponding lowest\footnote{Picking the lowest weight state is convenient when matching states with the Krylov basis.} weight state $\ket{\Lambda}$, annihilated by 
 any lowering ladder operator
 , as base $\ket{\psi_0}$ for a Lie group valued coherent state
 \begin{align}\label{eq:coherent_state_gen}
 |\Lambda(\xi)\rangle
 = 
 \exp\left(i\sum_{\alpha} \xi_\alpha E_{\alpha} +i\sum_{i}\xi_i H_i\right) |\Lambda\rangle\,
 \qquad
 \ket{\Lambda(0)}=\ket{\Lambda}\,,
 \end{align}
 The non-trivial individual states that make up the coherent state in \eqref{eq:coherent_state_gen} are precisely the descendant states generated by the action of the raising operators. In other words, the coherent state is being built up by adding contributions from increasing powers of the raising operators, which ``climb'' the weight lattice of the highest weight representation $\rho_\Lambda$ step by step.

Let us now illustrate how spread complexity is computed, reviewing two known examples of coherent states associated with the Heisenberg-Weyl and $\sl{2}$ Lie algebras. They play a central role in the following sections. Indeed, both are identified below as `atomic components' of spread complexity of more general Lie groups and in particular that of string states.

\subsubsection*{A simple example: HW-coherent state}\label{sec:HW}
The spread complexity for Heisenberg-Weyl coherent states has been computed already in \cite{Caputa:2021sib}, which we now review.

The starting point are the usual creation and annihilation operators of the harmonic oscillator $a^{\dagger}_B$ and $a_B$ with non-trivial commutator $[a_B,a_B^{\dagger}]= \mathds 1$.
The number operator $N_B=a_B^{\dagger}a_B$ acts on the $n$-th number state $\ket{n}$ by extracting the occupation number of bosonic particles, $N_{B} |{n}\rangle = n_{B} |{n}\rangle$. 
Creators and annihilators act on the number states via
\begin{align}
 a_B^{\dagger} |n\rangle=\sqrt{n_B+1}|n+1\rangle\,,\quad a_B|n\rangle =\sqrt{n_B}|n-1\rangle.
\end{align}
 A coherent state $\ket{\xi}$ of the bosonic oscillator is defined as eigenstate of the annihilation operator. Such are constructed by applying the displacement operator 
 \begin{equation}\label{HWdisplacer}
      D(\xi)=  \exp(\xi a_B^\dagger-\bar{\xi}a_B)
 \end{equation}
 to the ground state,
 \begin{eqnarray}
     |\xi\rangle
     =
     D(\xi)|0\rangle
     =
     e^{-|\alpha t|^2/2}\sum_{n=0}^\infty \frac{(i\alpha t)^n}{\sqrt{n!}}|n\rangle
     \label{bosonicCoherentState}\,,
 \end{eqnarray}
 where the time-dependence $\xi=i \alpha t$ is determined by the equations of motion of the semiclassical approximation. The probability of occupying the $n^{\rm th}$ Krylov state is 
 \begin{equation}\label{bosHWprob}
     p_n
     =
     |\braket{n}{\xi}|^2
     =
     e^{-|\xi|^2}\frac{|\xi|^{2n}}{n!}
 \end{equation}
 The complexity for the coherent state as, 
 \begin{equation}
    \mathcal{C}(t)
     =
     \hspace{-2pt}\bra{\xi}N\ket{\xi}
     =
     |\xi|^2
     =
     \alpha^2t^2\,,
     \label{FreeBosonCplx}
 \end{equation}
 where the number operator has been identified with the spread complexity operator, $\hat K=N$.

\subsubsection*{A further simple example: $\texorpdfstring{SL(2)}{SL2}$-coherent state}
Although the simplicity of the HW-coherent state is particularly appealing to illustrate the machinery of the spread complexity, it misses a property of most algebras we will consider in the what follows: semi-simplicity. Semisimple algebra are characterised by a particularly simple structure: in the Chevalley basis the algebra essentially breaks down into $n$ copies of $\mathfrak{sl}(2)$ knitted together by the particular root structure. The spread complexity for the $\mathfrak{sl}(2)$ was first computed in \cite{Balasubramanian:2022tpr}, which we now review for later reference.

The $\mathfrak{sl}(2)$-algebra is defined by three generators satisfying the commutation relations $[L_0, L_{\pm 1}] = \mp L_{\pm 1}\,, \; [L_1, L_{-1}] = 2L_0$. Taking the displacement operator of the coherent state to be
\begin{equation}\label{Sl2dDisplacer}
    D(\alpha,\gamma)=\exp\left[i(\alpha  (L_{-1} + L_1) + \gamma L_0)t\right]
\end{equation}
and identifying its exponent with the Hamiltonian. We pick a highest-weight irreducible representation of $\mathfrak{sl}(2,\mathbb C)$ labelled by $h$, then act on it with $D(\alpha,\gamma)$, the Krylov basis is 
\begin{equation}\label{sl2Kbasis}
    \ket{K_m}
    =
    \ket{h, h+m}
    =
    \frac{L_{-1}^m\ket{h}}{\norm{L_{-1}^m\ket{h}}},
    \qquad
    \norm{L_{-1}^m\ket{h}}^2
    =
    m!\frac{\Gamma(2h+m)}{\Gamma(2h)}\,,
\end{equation}
where the norm of a vector is denoted by $\norm{\ket{v}}=\sqrt{\braket{v}{v}}$. 
 The probability that a coherent state resides in the $m^{\rm th}$ basis vector is
 \begin{equation}\label{sl2Prob}
     p_m
     =
     \frac{\tanh^{2m}(\alpha)}{\cosh^{4h}(\alpha)}\frac{\Gamma(2h+m)}{m!\,\Gamma(2h)}
 \end{equation}
The resulting complexity can be computed to take the form \cite{Balasubramanian:2022tpr}
\begin{align}\label{Sl2Cplx}
    \mathcal C(t) 
    =
    \bra{\alpha,\gamma,h}L_0-h\ket{\alpha,\gamma,h}
    =
    \frac{2h}{1 - \frac{\gamma^2}{4\alpha^2}} \sinh^2 \left( \alpha t \sqrt{1 - \frac{\gamma^2}{4\alpha^2}} \right)\,.
\end{align}
Note that the spread complexity can be periodic or exponentially growing, depending on the values of the parameters $\alpha$ and $\gamma$. In particular, we do not wish to give them a particular interpretation.  
In fact, let us remark that at this point the particular chosen time dependence of the parameters, may seem ad-hoc; we address this matter in detail in subsection \ref{sec:semiclassics}.

\subsection{Super Lie group-valued coherent states and their complexity}\label{secSuperDisplacers}

Besides bosonic degrees of freedom, most systems, be it in particle physics, string theory or condensed matter, require fermionic degrees of freedom. This leads to the natural emergence of superalgebra symmetries. Before discussing how fermionic degrees affect spread complexity, we first review superalgebras and their supercoherent states.

Super Lie algebras generalise Lie algebras to include a $\mathbb{Z}_2$-gradation, called Grassmann parity. The gradation indicates the bosonic (even gradation) or fermionic (odd gradation) character of the generators, which is also reflected in Hilbert space
\begin{equation}
    \cH=\cH_\mathrm{even}\oplus\cH_\mathrm{odd}\,.
\end{equation}
Even elements close under the commutator, while odd elements produce an even element under the anticommutator. The commutator of an even element with an odd element is odd. Semiclassical superalgebras accordingly admit an analogue to the Cartan basis\footnote{See \cite{Frappat:1996pb} for a pedagogical review on Chevalley basis and root system for superalgebras.} and the previous discussion carries over apart from some critical changes which we now detail.

We thus consider coherent states whose displacement operator is superalgebra-valued, acting on a ground state $\ket{\Lambda}$ with definite Grassmann parity,
\begin{align}\label{superDisplacer}
|\Lambda(\xi,\zeta)\rangle = \exp\left(i\sum_{\alpha} \xi_\alpha E_{\alpha} +i\sum_A \zeta_A G_A+i\sum_{i}\xi_i H_i\right) |\Lambda\rangle\,,
\end{align}
where we denoted odd (or fermionic) generators by $G_A$. In addition, the corresponding coefficients $\zeta_A$ are now Grassmann-valued, so that this displacement operator is Grassmann even. Therefore, $\ket{\Psi(\xi,\zeta)}$ inherits is Grassmann parity from $\ket{\Lambda}$.

Let us discuss how to treat Grassmann numbers when calculating spread complexity with coherent states \eqref{superDisplacer}.
Grassmann numbers play a crucial role in formulating theories with fermionic statistics and extracting physical predictions. Indeed, all observables have even statistics. As we will see below, the presence of Grassmann parameters propagates, naively, all the way into physical quantities such as probabilities and complexities. Thus, we require a method to extract a bosonic number from elements in a Grassmann algebra, say $A=z+y\zeta_0+x\bar\zeta_0+w\bar\zeta_0\zeta_0$ where $z,y,x,w\in\C$, $\zeta_0$ is a Grassmann odd unit vector in the algebra and $\bzeta_0$ is its complex conjugate. We achieve this via Grassmann integration
\begin{equation}\label{scAverage}
    \gev{A}
    =
    \int d\zeta_0 d\bar\zeta_0\,e^{\bar\zeta_0\zeta_0}\,A
    =
    z+w\,,
\end{equation}
where the subscript stands for \quotes{semiclassical average}, which effectively condenses the purely fermionic degrees of freedom. This averaging protocol has the physically desirable properties of removing all fermionic statistics
\begin{align}\label{eq:semicl_av_details}
    \gev{ z }
    =
    z \qquad
    \gev{ \bar \zeta_0 \zeta_0 }
    =
    -\gev{  \zeta_0\bar \zeta_0 }
    =
    1\,,\quad 
    \gev{ \bar \zeta_0 }= \gev{\zeta_0 } =0\,.
\end{align}
This averaging is illustrated on page \ref{secGrassmannOddParameter} for the free fermion in order to extract probabilities and spread complexities. Concretely, the spread complexity is
\begin{equation}\label{SpreadCplxSusy}
    \cC(\xi,\chi,\Lambda)
    =
    \int d\zeta_0 d\bar\zeta_0\,e^{\bar\zeta_0\zeta_0}\bra{\Lambda(\xi,\chi)}K\ket{\Lambda(\xi,\chi)}\,,
\end{equation}
where $\chi$ is a fermionic label and $K$ the Krylov operator.

\subsection{Coherent state path integral and semiclassical approximation}\label{sec:semiclassics}
\begin{figure}[t]
    \centering
    \includegraphics[width=0.68\linewidth]{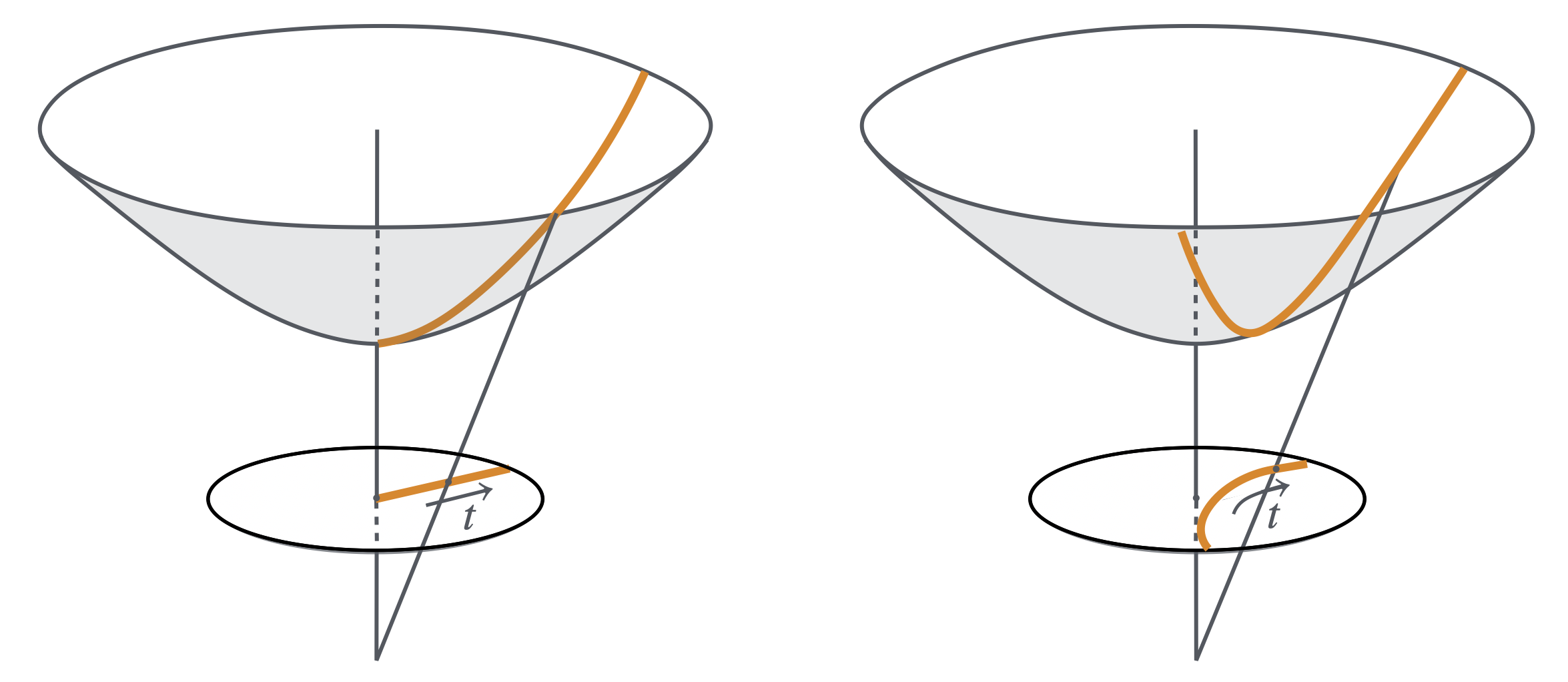}
    \caption{The time evolution in the spread complexity is fixed by solving the semi-classical path in phase space of the coherent state. On the left the path in phase space for the linear $SL(2,\mathbb R)$ Hamiltonian. Choosing a different Hamiltonian, for the same set of coherent states, leads to a different path in phase space. Such a hypothetical path is sketched on the right. Although described by the same set of coherent states, the resulting complexity is distinct due to a different time evolution.}
    \label{fig:hyperbolic_disc}
\end{figure}

We now turn to the question alluded to earlier: how the time-dependence of spread complexity is fixed for a given Hamiltonian evolution. Here we review how this is answered by fixing the phases space trajectory travelled by the initial state under the action of the Hamiltonian function. In addition, we clarify how Grassmann parameters, which are necessary when working with unitary supergroup displacement operators, are handled to extract physical probabilities and quantities.

Coherent-state path integrals were first introduced by Klauder in \cite{klauder1979path} (see also \cite{klauder1985coherent,Fradkin:2013sab} for pedagogical introduction and derivations). In this semiclassical approximation, one expands the action around the classical path by specialising to a coherent state representation of the degrees of freedom. The aim is to compute a transition amplitude $ \mathcal T$ between a coherent state $|\Psi(\xi_1)\rangle$ at an initial time $t_i$ and a second
coherent state $|\Psi(\xi_2)\rangle$ at time $t_f$ for a system governed by a Hamiltonian $H$, which is given by \cite{klauder1979path} 
\begin{align}
   \mathcal T=\int \mathcal D \Psi \,e^{iS(\Psi)}\,,
\end{align}
with the action given by 
\begin{align}\label{eq:semi_class_act}
    S=\int_{t_i}^{t_f} \mathrm d t\, L=\int_{t_i}^{t_f} \mathrm d t \Big(\langle \Psi|\partial_t |\Psi\rangle-\langle \Psi|H|\Psi\rangle\Big)\,.
\end{align}
 The semiclassical equations of motion follow from the usual Euler-Lagrange equations from the Lagrangian $L$ or by evaluating the Poisson brackets in the Hamiltonian picture, see section VI of \cite{Caputa:2021sib} for details. Averaging the quantum Hamiltonian $H$ over the coherent state basis $\mathsf H\equiv \langle \Psi|H|\Psi\rangle$ yields  a Hamilton function. Solving the equations of motion derived from the effective action in eq. \ref{eq:semi_class_act}, pins down the evolution of the coherent states in terms of their parameters, i.e. they determine $\xi=\xi(t)$ in eq. \eqref{eq:coherent_state_gen} where $t$ is the \quotes{time} for the Hamiltonian function $\mathsf H$.

Turning back to the $SL(2)$-coherent state reviewed in section \ref{sec:review_spread_C}, the phase space of the coherent state is described by the complex number parameter $\xi$  which in turn is a point in the hyperbolic disc $G/G_{\psi_0}\cong SL(2)/U(1)$, where we have denoted the lowest weight state by $\psi_0$. Writing  $\xi=\tanh(\rho/2)e^{i\phi}$, where $\rho\in\R_+$ and $\phi\in [0,2\pi)$ describe the radius and angle of the disc, respectively, the semiclassical equations of motion for the linear Hamiltonian 
\begin{align}\label{Sl2dDisplacer2}
    H_{SL(2,\mathbb{R})} = \xi( L_1 +  L_{-1})\,,
\end{align}
admits the solution $\phi=\pi/2$ and $\rho=2\alpha t$. In terms of a trajectory in the phase space of the coherent state, this corresponds to a path along a ``great circle'' of the hyperbolic disk at a fixed angle, see fig. \ref{fig:hyperbolic_disc}.

As already mentioned in the introduction, the linear Hamiltonian in e.g. \eqref{Sl2dDisplacer2} is but one possible system for which the $SL(2)$-coherent states describe the semiclassics. Notable examples, which we  discuss in the last section, are spin chains. These, generically more involved, Hamiltonians drive a different semiclassical path in the phase space of the pertinent coherent states. More concretely, they lead to a different functional dependence $\xi=\xi(t)$ and different complexities.

\section{Spread complexity with fermions}\label{sec:fermion_osc}

In this section we consider coherent states generated by the fermionic generalisation of the Heisenberg-Weyl algebra. With the ingredient introduced in the previous section, we consider the first step towards spread complexity of superstring states in the planar limit: computing spread complexity in fermionic coherent states.

\subsection{Fermionic Heisenberg-Weyl algebra}\label{sec:FermionicHW}

In view of introducing fermionic coherent states, 
we now consider oscillators
where the creation and annihilation operator satisfy the anti-commutation relations
\begin{align}
    \{a_F,a_F^\dagger\}=1\,,\quad \{a_F,a_F\}=\{a_F^\dagger,a_F^\dagger\}=0\,.
\end{align}
As is well known, the eigenstates of its number operator can only be 0 or 1:  $N_F|n\rangle=a_F^\dagger a_F|n\rangle=n|n\rangle$ where $n\in\{0,1\}$.
While for bosons, an $n$-particle state takes on the form $|n_B\rangle=\frac{1}{\sqrt{n!}}(a_B^\dagger)^n|0\rangle$.
For fermions, an $n$-particle state takes on the form
$|n_F\rangle = (a_F)_1^\dagger (a_F)_2^\dagger \cdots (a_F)_n^\dagger|0\rangle$.

The fermionic coherent states for the free fermionic oscillator or Clifford algebra is then the state generated from the displacement operator acting on the vacuum\footnote{ Note that sometimes the coherent state is defined as 
$|\theta\rangle =e^{-\theta a_F^\dagger}|0\rangle\,$,
and since $ a_F|0\rangle=0$ it follows that $a_F|\theta\rangle =\theta |\theta\rangle$.}
\begin{align}\label{eq:FHW_displ}
    |\xi\rangle =D(\xi)|0\rangle=e^{\xi a_F^\dagger - a_F\bar\xi}|0\rangle= \sum_n \frac{(\xi a_F^\dagger -a_F\bar\xi )^n}{n!}|0\rangle\,,
\end{align}
where we have defined the displacement operator $\displaystyle{D(\xi)=\exp(\xi a_F^\dagger - a_F\bar\xi)}$ and  $\xi$ is a complex (Grassmann even or odd) number and the bar denotes its complex conjugate. We will focus on each possibility in turn.

\subsubsection*{Grassmann even parameter} \label{secGrassmannEvenParameter}
Here, we consider the simplest case where, due to the statistics of the states, we lose the infinite nature of the usual semi-infinite chain seen in section \ref{sec:review_spread_C}.

The starting point is the coherent state and its displacement operator in eq. \eqref{eq:FHW_displ}. When $\xi$ is Grassmannian even, there is actually a nice closed way to sum this series since the fermionic operators are nilpotent. The result is 
\begin{align}
    D(\xi)=\cos |\xi|+\sin |\xi|(e^{i\phi}a_F^\dagger-e^{-i\phi}a_F) \,,\label{eq:fermionic_disp}
\end{align}
where $|\xi|=\sqrt{\xi\bar\xi}$ and $\phi$ is in $\xi=|\xi|e^{i\phi}$. 
Note that the coherent state for the fermionic oscillator coherent state is dramatically different from the bosonic one (not surprisingly) consisting of just two states rather than an infinite sum of states. 

\begin{figure}[t]
    \centering
\tikzset{every picture/.style={line width=0.75pt}} 
\begin{tikzpicture}[x=0.75pt,y=0.75pt,yscale=-1,xscale=1]

\draw   (26.77,20.4) .. controls (26.77,15.92) and (30.4,12.29) .. (34.89,12.29) .. controls (39.37,12.29) and (43,15.92) .. (43,20.4) .. controls (43,24.88) and (39.37,28.51) .. (34.89,28.51) .. controls (30.4,28.51) and (26.77,24.88) .. (26.77,20.4) -- cycle ;
\draw    (43,20.4) -- (99.77,20.4) ;
\draw   (99.77,20.4) .. controls (99.77,15.92) and (103.4,12.29) .. (107.89,12.29) .. controls (112.37,12.29) and (116,15.92) .. (116,20.4) .. controls (116,24.88) and (112.37,28.51) .. (107.89,28.51) .. controls (103.4,28.51) and (99.77,24.88) .. (99.77,20.4) -- cycle ;

\draw (10,36.4) node [anchor=north west][inner sep=0.75pt]    {$m=0$};
\draw (85,36.4) node [anchor=north west][inner sep=0.75pt]    {$m=1$};
\end{tikzpicture}
\caption{Spread complexity for bosonic states, for example the $\mathfrak{sl}(2,\mathbb R)$ coherent state, leads to an effective semi-infinite chain. For purely fermionic states, the chain becomes finites, reflecting the statistic of those states.}\label{fig:fermchain}
\end{figure}
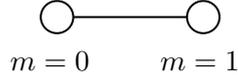

From this explicit form for the displacement operator the fermionic coherent states for the free fermionic oscillator becomes
\begin{align}\label{eq:single_mode_coherent state}
    \ket{\xi}_F 
    =
    D(\xi)\ket{0}
    =
    \cos |\xi||0\rangle+e^{i \phi}\sin |\xi||1\rangle \,.
\end{align}
In terms of spread complexity, we can identify in a similar way as in \cite{Caputa:2021sib} we can identify $\xi=i \alpha t$, where $t$ is time and $\alpha$ is a constant.
Note that this is a finite sum and implies that the Krylov space is spanned by just two basis elements: $|0\rangle$ and  $|1\rangle$, i.e. a fermion or no fermion. Any operator in this space is a finite linear combination
\begin{align}
    |\mathcal \psi(t)\rangle =\sum_{m=0}^1\varphi_n(t)| K_m\rangle= \varphi_0(t)|0\rangle + \varphi_1(t)|1\rangle\,.
\end{align}
This has to be contrasted with e.g. the infinite sum for the $\mathfrak{sl}(2,\mathbb R)$-coherent state in eq.  (41) of \cite{Caputa:2021sib}.
As a result, we no longer have a semi-infinite chain, see figure \ref{fig:simple_semi_chain}, but a chain consisting of just two nodes, depicted in figure \ref{fig:fermchain}.

Thus the complexity for the fermionic coherent state, is given by
\begin{equation}
   \mathcal{C}(t)
      =
   \bra{\psi(t)}N_F\ket{\psi(t)}
   =
   \sum_{m=0}^1m|\varphi_{m}(t)|^2=\sin(\alpha t)^2\,.
\end{equation}
As expected, the Krylov complexity, or expectation value for the position of the fermion in the `chain', is periodic due to the chain's finiteness. We point out that by picking $\ket{1}$ as ground state for the evolution, we find $\cC(t)=\cos^2(\alpha t)$ as spread complexity.

\subsubsection*{Grassmann odd parameter}\label{secGrassmannOddParameter}

We now turn to the case where the variable in the displacement operator in eq. \eqref{eq:fermionic_disp} is taken to be Grassmann odd and denote it by $\zeta$. We thus consider now the anti-hermitian\footnote{It is convenient to define the complex conjugation of multiple Grassmann variables as follows $\overline{(\zeta_1\zeta_2)}=\bar \zeta_2\bar \zeta_1$.}  displacement operator and coherent state \begin{align}\label{fermionCoherentState}
    \ket{\zeta}
    =
    D(\zeta)\ket{0}
    = e^{ a_F^\dagger\zeta - \bar \zeta a_F}  \ket{0}=
    \left(1- \frac{1}{2}\bar\zeta\zeta\right)\ket{0} +\zeta \ket{1} \,.
\end{align}
The displacement operator  can be easily shown to verify $D(\zeta)^\dagger=D(\zeta)^{-1}=D(-\zeta)$.
The wave functions associated to two states of the Hilbert space follow immediately,
\begin{equation}
    \varphi_0=\braket{0}{\zeta}=\left(1- \frac{1}{2}\bar\zeta\zeta\right), \qquad
    \varphi_1=\braket{1}{\zeta}=\zeta\,.
\end{equation}
We observe that their absolute values, $|\varphi_i|^2$, depend on the Grassmann variables $\zeta$. In Appendix \ref{app:FermionSymplectic} we show that $\zeta=i\alpha\zeta_0 t$, where $\zeta_0$ is again a unit Grassmann vector and $\alpha\in\R$. As discussed in subsection \ref{secSuperDisplacers}, the probabilities are then defined by averaging over the Grassmann parameters,
\begin{align}
    p_i
    &=
    \int d\zeta_0 d\bar\zeta_0 \,e^{\bar\zeta_0\zeta_0}|\varphi_i|^2
    =\begin{cases}
        &(1-\alpha^2 t^2)\,, \quad \text{for } i=0\,,\\
        & \alpha^2 t^2\,, \qquad\quad\;\;  \text{for } i=1\,,
    \end{cases}
\end{align}
where the subscript $i=0,1$ indicates the ground state or the one-fermion state, respectively.
These probabilities are bosonic numbers, which add up to one, as desired. Note however that the evolution is ill-defined once $|\alpha t|>1$, since the probabilities become unphysical in this case.

The complexity is computed similarly by semiclassical averaging
\begin{align}\label{FermionCplxty}
    \cC
    =
    \int d\zeta_0 d\bar\zeta_0 \,e^{\bar\zeta_0\zeta_0}\bra{\zeta}N_F\ket{\zeta}
    =
   (\alpha t)^2\gev{ \bzeta_0\zeta_0}
    =
    \alpha^2 t^2\,.
\end{align}
Observe that this result is the same as \eqref{FreeBosonCplx}, except for the important fact that $|\alpha t|\leq1$. When $|\alpha t|=1$ the system has finished its evolution from the initial state $\ket{0}$ to the other remaining state $\ket{1}$ and remains there. We see that the spread complexity is bounded. This behaviour, has already been observed for bosonic systems with finite-dimensional Krylov subspace $\cK$. This is precisely what the bosonic character of the total displacer here mirrors.

\subsection{Superposition of fermionic HW algebras}
\label{subsec:ferm_mf}

We extend our analysis of the complexity associated with single-fermionic coherent states to include multi-fermionic coherent states. This extension aims to investigate the impact of a higher-dimensional Hilbert space on complexity, when only fermionic states are considered. Additionally, multi-fermionic coherent states naturally arise in the path-integral formulation used to quantise fermionic fields. In the case of a fermionic system with chiral fermions in a finite size interval $[-L/2,L/2]$, the momentum $k$ satisfies \cite{Ge_2019} $  k=\frac{2\pi n}{L}$ for$~n\in \mathbb{Z}$. For a single fermion, the mode decomposition is given as
\begin{align}
    \psi(x)= \left(\frac{2\pi}{L}\right)^{1/2}\sum_{n=-\infty}^\infty e^{-i \frac{2\pi n}{L}x}(a_F)_n,~~(a_F)_n=(2\pi L)^{-1/2}\int_{-L/2}^{L/2}dx~e^{i\frac{2\pi n}{L}x}\psi(x)\,.
\end{align}
These fermionic creation and annihilation operators $(a_F)_k^\dagger$ and $(a_F)_k$ for momenta modes $k$ satisfy the following anticommutation relation, $\{(a_F)_k,(a_F)^\dagger_l \}=\delta_{kl}$.

\subsubsection*{Grassmann even parameters}
A class of $N$-mode fermionic coherent states in terms of these operators are defined as \cite{Fan:1998cb}, 
\begin{align}
    |\{\xi_k\}\rangle=\bigotimes_{k=1}^N \left(\cos{|\xi_k|}|0\rangle_k+e^{i\phi_k}\sin{|\xi_k|}|1\rangle_k\right)\,,
\end{align}
here the variables $\xi_k$ are taken to be complex Grassmann even numbers. In this definition of the $N$-mode fermionic coherent state, the displacement operator is the product of $N$ single fermionic displacement operators~\eqref{eq:fermionic_disp} corresponding to individual momenta modes.

Next, to elevate the multi-mode displacement operator to a time evolution operator, we take $\xi_k = i\alpha_k t$ as before, where $\alpha_k$ is a mode-dependent displacement parameter. Starting from the all-mode vacuum state $|0\rangle$, the state at time $t$ is given by:  
\begin{align}
\begin{split}
  |\{\alpha_k\},t\rangle &= \bigotimes_k \cos(|\alpha_k t|)\left(|0\rangle_k + e^{i\phi_k} \tan(|\alpha_k t|) |1\rangle_k\right) \\ 
  &= \bigotimes_k \lambda_k\left(|0\rangle_k + \eta_k|1\rangle_k\right) \,,\label{eq:multi_mode_coh_st}
\end{split}
\end{align}
where we define $\lambda_k = \cos(|\alpha_k t|)$ and $\eta_k = e^{i\phi_k} \tan(|\alpha_k t|)$ for simplicity in the calculations.  

The Krylov basis vectors in this case are constructed starting from the all-mode vacuum $|0\rangle$ via the action of $c_k^\dagger$ operators for different values of $k$. The total number of fermions in a given Krylov basis vector quantifies how `far' it has evolved from $|0\rangle$. For instance, both $|100\ldots\rangle$ and $|0100\ldots\rangle$ are one step away from the initial state $|0\rangle$. Similarly, $|101\ldots\rangle$ and $|0111\ldots\rangle$ represent two and three steps, respectively. This evolution can be understood through the $N$-fermion displacement operator, $\bigotimes_{i=1}^N\displaystyle{D(\xi_i)}$ where $\displaystyle{D(\xi_i)}=\exp(\xi_i a_F^\dagger - \bar{\xi}_i a_F)$, applied to the $N$-fermion vacuum state.  

For the simplest case where $\xi_i = \xi$ for all $i$, the Krylov basis forms a uniform linear superposition of fermion number states with a fixed number of fermions. In more generic cases, the Krylov basis represents a non-uniform superposition, which is computationally challenging to determine for arbitrary $\xi_i$ values. 

Nevertheless, for general values of $\xi_i$, we can compute the complexity of spreading of the time-evolved state in the Fock basis of the system, weighting the probabilities by the total number of fermions present in each state. It is important to note that the Fock basis or its uniform superposition is not equivalent to the Krylov basis of the system, except for the specific case $\xi_i = \xi$ for all $i$. However, this complexity of spreading in the Fock basis is an interesting quantity in its own right and serves as an upper bound on the spread complexity of the system \cite{Balasubramanian:2022tpr}, the reason being simply that the Fock basis is much larger than the Krylov basis can be. For a given set of occupation numbers $\{n_k\}$, we label each Krylov basis vector as $|\{n_k\},N\rangle$, where $N$ is the total number of fermions in the state.

For the single-mode case, the multi-fermion coherent state in eq. \eqref{eq:multi_mode_coh_st} reduces to eq. \eqref{eq:single_mode_coherent state} and complexity becomes, $\mathcal C= \lambda^2|\eta|^2=\sin{(\alpha t)}^2$. Going one step further, in the notation used above, the double-mode coherent state is given by
\begin{equation}
 |\alpha_{1},\alpha_{2},t\rangle= (\lambda_{1}\lambda_{2})\left(|00\rangle+\eta_{1}|1_{1}0\rangle+\eta_{2}|01_{2}\rangle+\eta_{1}\eta_{2}|1_{1}1_{2}\rangle\right)\,
\end{equation}
The (normalised) wavefunctions, for say $\ell$ out of $N$ fermions, are thus of the form
\begin{align}
    \varphi_{\ell,\{\epsilon\}}(t)=\left(\prod_{i=1}^\ell \lambda_k\right)\eta_1^{\varepsilon_{1,\ell}}\cdots \eta_N^{\varepsilon_{N,\ell}}\,,
\end{align}
where  we have defined $\{\epsilon\}\equiv\{\epsilon_{1,\ell},\epsilon_{2,\ell},\dots,\epsilon_{N,\ell}\}$, where each $\epsilon_{i,\ell}\in\{0,1\}$ and tracks the occupancy of a fermionic state in the $i^\mathrm{th}$ entry of tensor product state. Note that $\sum_i\epsilon_{i,\ell}=\ell$. The associated probabilities in the Fock basis of the system are simply
\begin{align}
    p_{\ell,\{\epsilon\}}(t)=\left(\prod_{i=1}^\ell \lambda_i^2\right)\eta_1^{2\varepsilon_{1,\ell}}\cdots \eta_N^{2\varepsilon_{N,\ell}}\,.
\end{align}
Their complexity of spreading for $\ell=2$ in the Fock basis is 
\begin{gather}
\begin{aligned}
    \mathcal{C}_F(t)&=(\lambda_{1}\lambda_{2})^2(|\eta_{1}|^2+|\eta_{2}|^2+2|\eta_{1}\eta_{2}|^2)\\
    &=
    1-\cos\left[(\alpha_{1}-\alpha_{2})t\right]\cos\left[(\alpha_{1}+\alpha_{2})t\right]\,.
\end{aligned}
\end{gather}
This is oscillatory behaviour but depending on the relative values of $\alpha_{k_1}$ and $\alpha_{k_2}$, the behaviour can be widely different. Following this prescription, for an $n$-mode fermionic coherent state, where we consider $n$ different momenta modes for the fermions, we obtain the complexity in the Fock basis as
\begin{align}
    \nonumber &\mathcal{C}_F(t)=\left(\sum_{i=1}^N|\eta_{i}|^2+\sum_{\substack{i,j=1\\ (i>j)}}^N2|\eta_{i}\eta_{j}|^2+...+\sum_{\substack{i_1,i_2,...i_N=1\\ (i_1>i_2>...>i_N)}}^NN|\prod_{j=1}^N\eta_{i_j}|^2\right)(\prod_{i=1}^N\lambda_{i})^2\\\nonumber&=\bigg(\sum_{i=1}^N|\tan{|\alpha_{i}t|}|^2+\sum_{\substack{i_1,i_2=1\\ (i_1>i_2)}}^N2|\prod_{j=1}^2\tan{|\alpha_{i_j}t|}|^2+...+\sum_{\substack{i_1,...,i_N=1\\ (i_1>...>i_N)}}^NN|\prod_{j=1}^N\tan{|\alpha_{i_j}t|}|^2\bigg)(\prod_{i=1}^N\cos{|\alpha_{i}t|})^2\\&=c_1(t)+c_2(t)+...+c_N(t)\,.\label{eq:total_MF_CF}
\end{align}
In the final line, we define the contribution of a sector with fixed particle number $m$ to the total complexity of spreading in the Fock basis
\begin{equation}
    c_m(t)=\sum_{\substack{i_1,...,i_m=1\\ (i_1>...>i_m)}}^Nm|\prod_{l=1}^m\tan{|\alpha_{l}t|}|^2(\prod_{i=1}^N\cos{|\alpha_{i}t|})^2\,.
\end{equation}
  The final form of \eqref{eq:total_MF_CF} cannot be simplified further for a general set of $\{\alpha_{k}\}$ coefficients. One possible simplification is if all the $\alpha_{k}=\alpha$ for all values of $k$. In this case, the last form of complexity simplifies completely and becomes 
\begin{equation}
    \mathcal{C}_F(t)=2^{N-2} N^2 (N+1) \sin ^2(\alpha t)\,. \label{eq:com_f}
\end{equation}
 In this case, complexity contributions from all the different components have the same frequency with time and they oscillate together. For this choice of parameters, the corresponding spread complexity is just a uniform linear of the fermion number states with the same number of total occupancy, and that differs from \eqref{eq:com_f} by a constant factor,

 \begin{equation}
     \mathcal{C}_F(t)=\frac{1}{6} N^2 (N+1) (2 N+1) \sin ^2(t)\,.
 \end{equation}

\begin{figure}[t]
     \centering
     \begin{subfigure}[b]{0.49\textwidth}
         \centering
         \includegraphics[width=\textwidth]{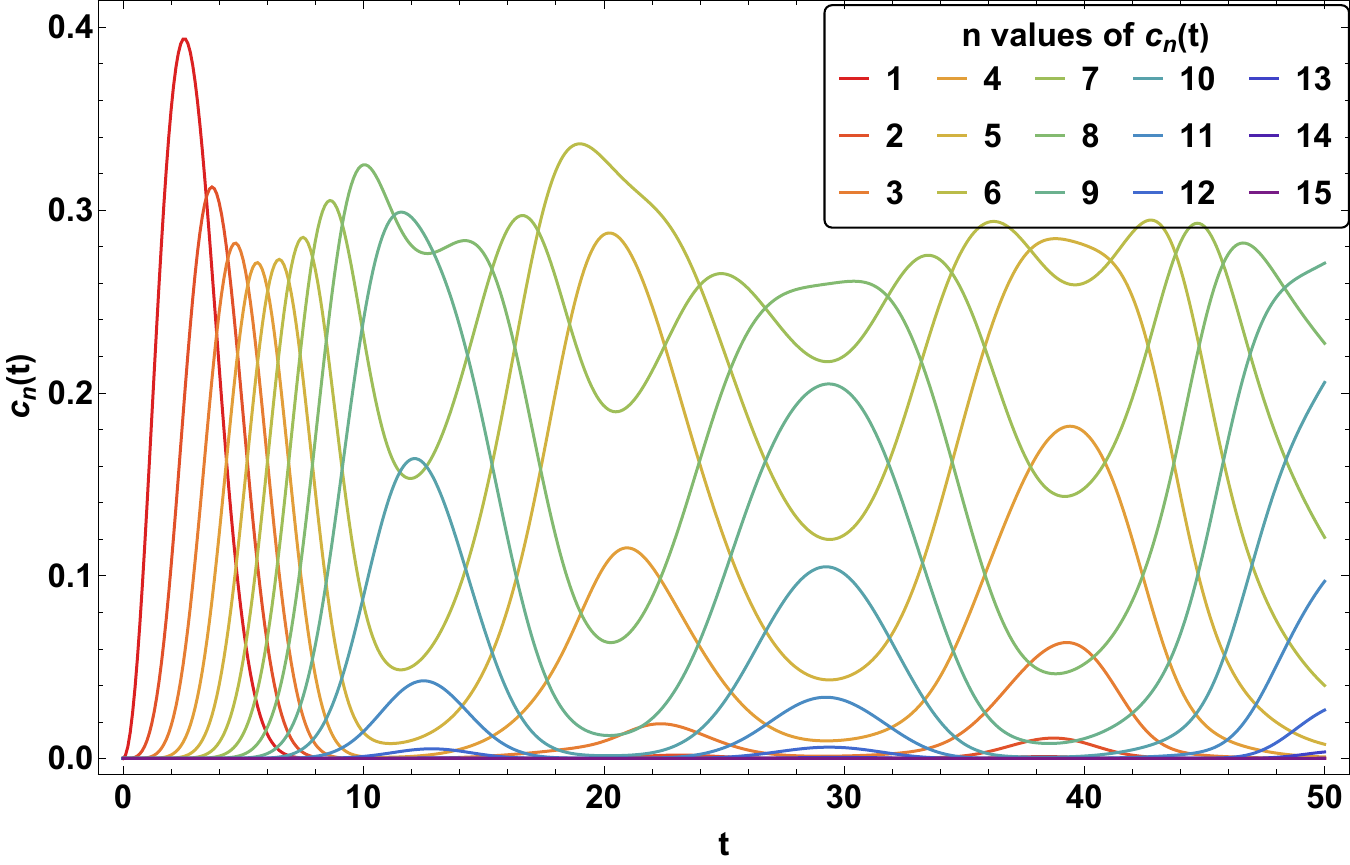}
         \label{fig:cn15}
     \end{subfigure}
     \hfill
     \begin{subfigure}[b]{0.49\textwidth}
         \centering
         \includegraphics[width=\textwidth]{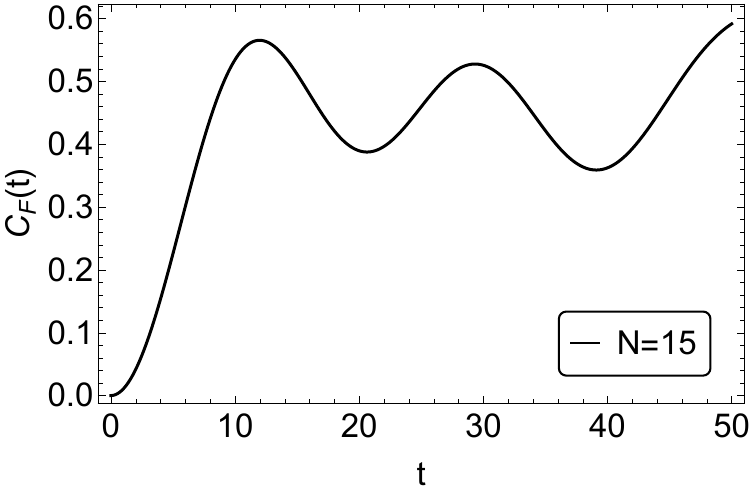}
         \label{fig:ck15}
     \end{subfigure}
     \vspace{-17pt}
        \caption{The left panel shows the plot of complexity contributions of spreading in the Fock basis for fixed total fermion number states normalised by the dimension of the Hilbert space. In the left panel, warmer colours indicate contributions from states with lower fermion numbers, while colder colours correspond to contributions from states with higher fermion numbers. At early times, the dynamics is dominated by states with smaller fermion numbers, whereas at later times, states with larger fermion numbers become dominant. The total complexity, shown on the right, is the envelope obtained by summing all contributions. On the right, the total normalised complexity of spreading in the Fock basis for the overall system for a 15-mode fermionic coherent state where the coefficients $\{\alpha_{k}\}$ are randomly chosen in the range [0, 1].  This illustrates how the total complexity of the system in the Fock basis arises as a cumulative contribution from individual fermion number states.
 }
        \label{fig:multi-mode_fermion_complexity_15}
\end{figure}

A more general and physically interesting case arises when the coefficients $\{\alpha_k\}$ are all distinct and randomly chosen from a given interval. This corresponds to displacing the vacuum state differently for each momentum mode. For this setup, the complexity of spreading in the Fock basis normalised by the Hilbert space dimension is illustrated in Fig.~\ref{fig:multi-mode_fermion_complexity_15}. In this scenario, we observe that the spread complexity of the collection of fermions does not exhibit oscillatory behaviour, although individual fixed-fermion-number components display oscillations and distinct peaks. As shown in the left panel of Fig.~\ref{fig:multi-mode_fermion_complexity_15}, the complexity contributions from states with smaller total fermion numbers dominate at early times, while those from states with larger total fermion numbers become significant at later stages of evolution. The total spread complexity thus represents a collective behaviour arising from the contributions of the spread complexity associated with states of different total fermion number, as depicted in the right panel of Fig.~\ref{fig:multi-mode_fermion_complexity_15}. It is worth mentioning that the technical details discussed in this section will play a crucial role in evaluating the spread complexity of rotating strings in purely fermionic subsector of the $PSU(1|1)$.

\subsubsection*{Grassmann odd parameters} 

 For products of displacement operators with Grassmann odd variables, i.e. of type \eqref{fermionCoherentState},
 \begin{align}
     D(\vec{\zeta}\,)\ket{0}
    =
    \prod_{k=1}^N
    e^{ (a_F)_k^\dagger\zeta_k - \bzeta_k (a_F)_k}  \ket{0}\,,
 \end{align}
 with anticommuting modes $\{(a_F)_k,(a_F)^\dagger_l \}=\delta_{kl}$ the situation is much simpler. Indeed, due to Grassmann oddness of the $\zeta_k$, the individual displacement operators commute. Taking the complexity operator $N_F=\sum_{k=1}^N(N_F)_k$, the total complexity is $\cC=\sum_{k=1}^NC_k(t)$, where each individual mode contributes a complexity $C_k(t)=\alpha_k^2t^2$, cf. eq. \eqref{FermionCplxty}.

\section{Spread complexity with bosons and fermions}\label{sec:superC}

\begin{figure}[t]
    \centering
    \includegraphics[width=0.33\linewidth]{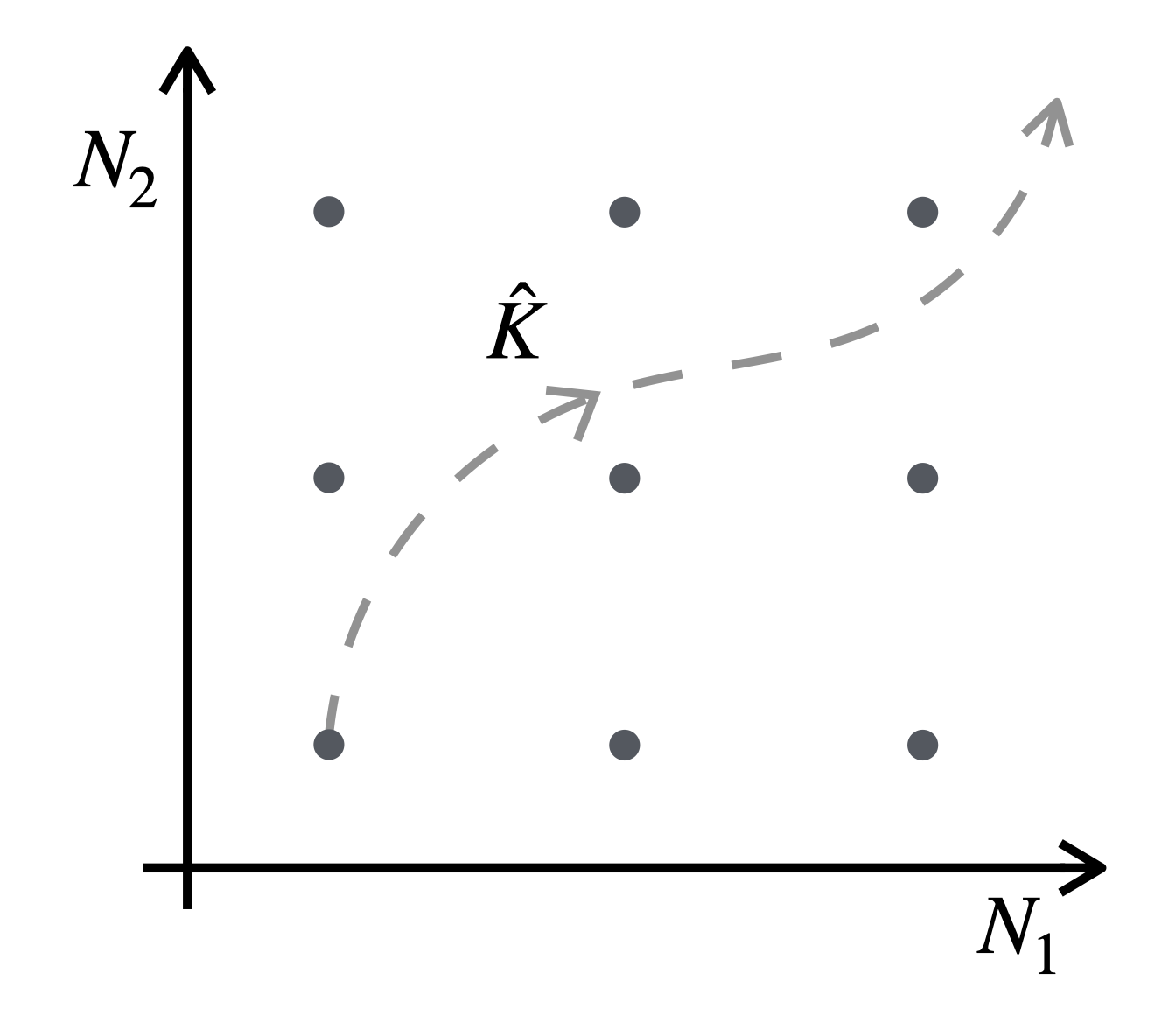}
    \caption{The lattice points represent a basis of the Hilbert space $\cH$, and a locus is read out by position operators $\vec{X}$. An initial vector $\ket{\psi_0}$ can be any superposition of these lattice points. Picking $\ket{\psi_0}$ to be the basis vector in the lower left corner, say, the evolution driven by a displacement operator $D$ leads through superpositions of basis vectors, i.e the \textit{Krylov path}  (dashed line). This path is an embedding of the Krylov chain into the total Hilbert space $\cK\hookrightarrow\cH$.}
    \label{fig:krylov_path}
\end{figure}

In this section we proceed to extend the notion of spread complexity to superalgebras. To begin, we treat the supersymmetric harmonic oscillator, see subsection \ref{sec:coh_superHW}. While simple, it already features all crucial aspects appearing when studying systems described by both bosonic and fermionic generators. These new features are however more pronounced in more involved superalgebras, prompting us to turn to the superalgebra $\osp{2|1}$ in a second step; see subsection \ref{sec:OSP}. These novelties are
\begin{itemize}
    \item The initial state $\ket{\psi_0}$ is now free to spread through a higher-dimensional lattice instead of a chain. When viewed through the lens of the superalgebra, this lattice is the weight lattice corresponding to the Hilbert space $\cH$ of a representation of the superalgebra. Therefore, the dimension of the lattice $r$ equals the rank of the algebra. In this section, the weight lattices take the shape of several semi-infinite ladders, so that $r=2$.
    \item When driven by a Hamiltonian, the state $\ket{\psi_0}$ follows its Krylov chain. Viewed from the perspective of the weight lattice, the Krylov chain describes a one-dimensional \textit{Krylov path} or \textit{spread path} through said lattice. This is a vector space embedding $\cK\hookrightarrow\cH$. We stress that this path is always one-dimensional, though exploring Hilbert space through linear combination of Fock space states, see figure \ref{fig:krylov_path}.
    \item The Krylov subspace $\cK$ $\subseteq\cH$ is generated by a single creation operator, usually a linear combination of several superalgebra generators. It naturally singles out an annihilator and a spread complexity operator. Together these three operators furnish a \textit{dynamical algebra} of rank one dictating the spread of $\ket{\psi_0}$ through $\cK\subseteq\cH$. 
\end{itemize}

In this novel situation we may ask how $\ket{\psi_0}$ spreads through the lattice, thereby probing the form of the embedding $\cK\hookrightarrow\cH$. In analogy with spread complexity, where the Krylov operator is the position operator on the Krylov chain, we define position operators $\vec{X}=(X_1,\dots,X_{r})$ on the lattice. Their expectation values measure the spread, or the average position, of a coherent state's Krylov path with respect to the directions of the lattice, prompting us to define \textit{lattice complexities}
\begin{equation}\label{SpreadCplxL}
    \vec{\cC}{}^L(\xi,\chi,\Lambda)
    =
    \left|\int d\zeta_0 d\bar\zeta_0\,e^{\bar\zeta_0\zeta_0}\bra{\Lambda(\xi,\chi)}\vec{X}\ket{\Lambda(\xi,\chi)}\right|\,.
\end{equation}
Note this is an $r$-dimensional vector of complexities. Furthermore, an absolute value is imposed to counter the orientation of the lattice, which may lead to a negative expectation value. For simplicity, we restrict to commuting position operators, $[X_i,X_j]=0$. Furthermore, in order for the lattice complexities to satisfy $C(0,0,\Lambda)=0$, we choose position operators satisfying $\vec{X}\ket{\Lambda}=0$. This requirement places the initial state $\ket{\Lambda}$ at the origin of the lattice.
\subsection{Super-Heisenberg--Weyl algebra}\label{sec:coh_superHW}

\begin{figure}[t]
    \centering
    \includegraphics[width=1\linewidth]{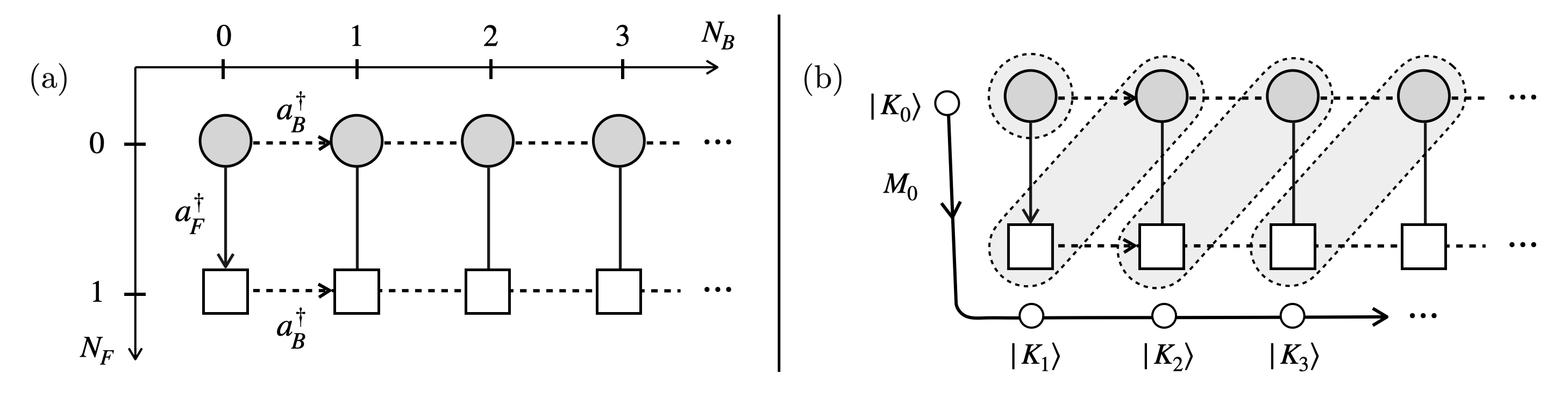}
    \caption{Left: The basis vectors $\ket{n,\nu}$ of $\cH$ construct a ladder with position operators $N_B,N_F$. Bosonic basis vectors are depicted as circles and fermionic ones as squares. Right: Picking $\ket{\psi_0}=\ket{0,0}$, the Krylov basis of an evolution driven by \eqref{superHWdisplacer} is a bosonic Fock space, whose occupation basis is a linear superposition of bosonic and fermionic basis vectors. The $n^{\rm th}$ Krylov basis vector is a superposition of the $n^{\rm th}$ fermionic and $n+1^{\rm st}$ bosonic lattice point.}
    \label{fig_ferm_HW}
\end{figure}

We consider the supersymmetric extension of the Heisenberg-Weyl algebra given by the annihilation and creation operators $a_B,a_B^\dagger,a_F,a_F^\dagger$ satisfying the (anti-)commutation relations $[a_B,a_B^\dagger]=\mathds 1\,,$ and $ \{a_F,a_F^\dagger\}=\mathds 1$.
These operators act on a (product) super Hilbert space $\mathcal H=\mathcal H_B\otimes \mathcal H_F$ factors, with elements which we denote by $|n,\nu\rangle\equiv |n\rangle \otimes |\nu\rangle$, where the first entry indicates the bosonic occupancy $n=0,1,2,\dots$ and the fermionic $\nu=0,1$. This infinite-dimensional  Hilbert space is spanned by $|n,\nu\rangle =(a_B^\dagger)^n(a_F^\dagger)^\nu |n,\nu\rangle/\sqrt{n!}$
and forms an orthonormal basis, which forms the lattice depicted on the left of figure \ref{fig_ferm_HW}. 
We call the states with $\nu=0$ bosonic and those with $\nu=1$ fermionic. Annihilators and creators move us through the lattice,
\begin{align}
	a_B|n,\nu\rangle
    &=
    \sqrt{n}|n-1,\nu\rangle\,,
    \quad
    a_B^\dagger|n,\nu\rangle
    =
    \sqrt{n+1}|n+1,\nu\rangle\,\\
	a_F|n,\nu\rangle
    &=
    \delta_{\nu 1}|n,0\rangle\,,
    \qquad\quad
    a_F^\dagger|n,\nu\rangle=\delta_{\nu 0}|n,1\rangle\,,
\end{align}
with $|0,0\rangle$ as ground state of the super Hilbert space. 

\noindent
For a unitary representation  of the supergroup, a coherent state is easily constructed using the displacement operator 
\begin{align}\label{superHWdisplacer}
D(\xi,\zeta
)= \exp(\xi a_B^\dagger -\bar \xi a_B +\zeta a_F^\dagger + \bar\zeta a_F).
\end{align}
Clearly, due to the commutation relations, this displacement operator neatly factorises into one displacement operator for bosonic and one for the fermionic degrees of freedom. Following this observation, we now adapt the procedures on page \pageref{sec:HW} and page \pageref{secGrassmannOddParameter} to construct a coherent state and expand it in the lattice basis $\ket{n,\nu}$,
\begin{align}\label{eq:sHW_coh_state}
	|\xi,\zeta\rangle 
    =
    D(\xi,\zeta)\ket{0,0}
    =
 \exp(-|\xi|^2/2)\sum_{n=0}^\infty\frac{\xi^n}{\sqrt{n!}}\Big((1-\tfrac{1}{2}\bar\zeta\zeta)|n,0\rangle+\zeta |n,1\rangle\Big)\,.
\end{align}
When $\zeta=0$, this state reduces to the bosonic Heisenberg-Weyl algebra coherent state in eq. \eqref{bosonicCoherentState} and for $\xi=0$, it reduces to \eqref{fermionCoherentState}. 

It is tempting to declare this lattice as the Krylov space $\cK$. This is, however, premature. Indeed, the basis $\ket{n,\nu}$ is \textit{not the Krylov basis} as seen by running the first iteration of the Lanczos algorithm: We see that $\ket{K_1}\propto (\xi a_B^\dagger+\zeta a_F^\dagger)\ket{0,0}$, which is a linear combination of points on the lattice in figure \ref{fig_ferm_HW}; the Krylov basis is introduced below. It is important to realise that the coherent state \eqref{eq:sHW_coh_state} knows about the evolution driven by $\xi a_B^\dagger -\bar \xi a_B +\zeta a_F^\dagger + \bar\zeta a_F$ regardless of which basis it expressed in. As explained above, working the lattice basis $\ket{n,\nu}$ allows us to probe how the Krylov chain is embedded into the lattice, i.e. it informs us about the Krylov path.

Following the standard prescription, wave functions are read off from \eqref{eq:sHW_coh_state}, 
\begin{align}
	\varphi_n^{\nu=0}(\xi,\zeta)
 &=
 \braket{n,0}{\xi,\zeta}
 =
\frac{\xi^n e^{-\frac{|\xi|^2}{2}}}{\sqrt{n!}}(1-\tfrac{1}{2}\bar\zeta\zeta)\,,
\qquad 
 \varphi_n^{\nu=1}(\xi,\zeta)
 =
 \braket{n,1}{\xi,\zeta}
 =
\frac{\xi^n e^{-\frac{|\xi|^2}{2}}}{\sqrt{n!}}\zeta\,.
\end{align}
These are \textit{not the Krylov wavefunctions} since they are associated with the individual lattice points in figure \ref{fig_ferm_HW} rather than points on the chain. To proceed, we bear in mind that $\xi=\xi(t)$, where the time parameter $t$ controlling the spread through the Krylov chain is discussed below; similarly we have $\zeta=f(t)\zeta_0$, with $\zeta_0$ a unit Grassmann odd element and $f(t)$ a c-valued function of time.

As discussed in subsection \ref{secSuperDisplacers}, the probabilities are conventionally found by $p_n=|\varphi_n|^2$, require semiclassical averaging over the Grassmann parameters,
\begin{align}
    p_n^\nu(t)
    &=
    \int d\zeta d\bar\zeta\, e^{\bar\zeta_0\zeta_0}|\varphi_n^\nu|^2
    = \begin{cases}
        & \frac{|\xi(t)|^{2n}}{n!}e^{-|\xi(t)|^2}(1-|f(t)|^2)
    \,, \quad\;\, \text{for } \nu=0\,.\\
     & \frac{|\xi(t)|^{2n}}{n!}e^{-|\xi(t)|^{2}}|f(t)|^2\,, \qquad \qquad \text{for } \nu=1\,.
    \end{cases}
\end{align}
As above, for these expressions to be probabilities, $0\leq|f(t)|^2\leq1$ must be respected. These $p_n^\nu$ indicate the probability that the state $\ket{\psi_0}$ sits at the basis vector $\ket{n,\nu}$ during the course of its evolution driven by \eqref{superHWdisplacer}. In geometric terms, this is the probability that the Krylov path passes through $\ket{n,\nu}$ at time $t$.  

The position operator $X$ measuring spread of the Krylov path along the horizontal direction in figure \ref{fig_ferm_HW} is $N_B$, whereas for the vertical direction it is $N_F$. Their respective lattice complexities \eqref{SpreadCplxL} quantify the spread induced by \eqref{superHWdisplacer} through the lattice spanned by the basis $\ket{n,\nu}$,
\begin{align}
   \cC^L_i(t,s)
    =
    \int d\zeta_0 d\bar\zeta_0\, e^{\bar\zeta_0\zeta_0}\bra{\xi,\theta}N_i\ket{\xi,\theta}
    =
    \begin{cases}
        &  |\xi(t)|^2\,, \quad \text{for } i=B\,,\\
        & |f(t)|^2\,, \quad \text{for } i=F\,. 
    \end{cases}
\end{align}
While the $N_i$ are Krylov operators for the displacement operators in \eqref{HWdisplacer} and in \eqref{eq:FHW_displ}, respectively, here they are not. Here they do not quantify spread through the Krylov chain itself, but indicate the spread of the chain relative to the lattice directions.

\subsubsection*{The Krylov basis and its complexity} 
We now turn to the spread complexity through the Krylov chain.
In order to find the Krylov basis, note that by using the \textit{bosonic} Heisenberg-Weyl operators
\begin{align}
    A=\frac{\xi a_B^\dagger+\zeta a_F^\dagger}{\sqrt{\bxi\xi+\bzeta\zeta}},
    \qquad
    A^\dagger=\frac{\bxi a_B-\bzeta a_F}{\sqrt{\bxi\xi+\bzeta\zeta}},
    \qquad
    [A,A^\dagger]=\id\,,
\end{align}
the displacement operator \eqref{superHWdisplacer} is recast as
\begin{equation}\label{HWdisplacerA}
    D(\xi,\zeta)
    =
    \exp\left[\sqrt{\bxi\xi+\bzeta\zeta}\left(A^\dagger-A\right)\right]\,.
\end{equation}
This is precisely the bosonic HW displacement operator \eqref{HWdisplacer} with $\xi=\bxi\to\sqrt{\bxi\xi+\bzeta\zeta}$. In other words, the Krylov basis $\ket{K_k}$ for the super-HW coherent states coincides with the Fock space of bosonic HW-states for the mode $A$. Note that the individual Krylov basis elements are superpositions of the ladder basis vectors $\ket{n,\nu}$. Hence, the Krylov chain lies along a \textit{Krylov path} through the super-HW lattice in the left panel of figure \ref{fig_ferm_HW}, which is parametrised by $(\xi,\chi)$.

Given this insight, every aspect of the spread complexity follows in analogy with the discussion on page \pageref{sec:HW}. The probabilities associated with each Krylov basis vector are 
\begin{equation}\label{eq:probs_SHW_KrylovB}
    p_k
    =
    \gev{e^{-(\bxi\xi+\bzeta\zeta)}\frac{\left(\bxi\xi+\bzeta\zeta\right)^k}{k!}}
    =
    e^{-|\xi(t)|^2}\frac{|\xi(t)|^{2k}}{k!}\left(1+|f(t)|^2\left(\frac{k}{|\xi(t)|^2}-1\right)\right)\,.
\end{equation}
The complexity is then  readily adapted from \eqref{FreeBosonCplx}
\begin{equation}
    \cC_\mathrm{HW}(t)
    =
    \gev{\bxi\xi+\bzeta\zeta}
    =
    |\xi(t)|^2+|f(t)|^2
    =
    (\alpha t)^2\,,
\end{equation}
where $|f(t)|^2\leq 1$ as above and $\alpha\in\R$. The functions $\xi,f$ need to be chosen such that the spread complexity $\cC_{HW}\propto t^2$ and $\xi(0)=f(0)=0$. An example is $|\xi|^2=(\alpha t)^2-\sin^2(\alpha t)$ and $|f(t)|^2=\sin^2(\alpha t)$. 

Summarising this section, we stress that it is possible to map the evolution driven by \eqref{HWdisplacer} containing bosonic and fermionic degrees of freedom into an effective bosonic HW system studied on page \pageref{sec:HW}. The corresponding Krylov chain follows a path through the lattice spanned by the basis vectors of $\cH=\cH_B\otimes\cH_F$. In the following, we turn to a more complicated example, where we re-encounter this structure.

\subsection{\texorpdfstring{OSp$(2|1)$}{Osp21} or \texorpdfstring{$\mathcal N=1$}{N1} \texorpdfstring{SL$(2)$}{Sl2}}\label{sec:OSP}
The simplest supersymmetric extension of $\mathfrak{sl}(2)$ is $\mathfrak{osp}(2|1)$-algebra. Besides the $\mathfrak{sl}(2)$ generators $L_0,\,L_{\pm}$ it contains two fermionic generators $G_{\pm 1/2}$ satisfying the graded commutation relations 
\begin{subequations}\label{ospAlgebra}
 \begin{align}
 [L_{+1},L_{-1}]&=2L_0\,,\qquad\qquad \quad [L_{\pm 1},L_{0}]=\pm L_{\pm1}\,,\\
 [L_{\pm},G_{\mp 1/2}]&=\pm G_{\pm 1/2}\,,\quad\, \{G_{1/2},G_{-1/2}\}=2L_{0}\,,\\
 \{G_{\pm 1/2},G_{\pm 1/2}\}&=2L_{\pm1}\,,\qquad \qquad[L_0,G_{\pm 1/2}]=\mp\frac{1}{2} G_{\pm 1/2}\,.
 \end{align}
 \end{subequations}
We also require an analogue of the fermion number operator, which counts the number of applications of the supercharge. This charge is called $J$ in this text and has the desired property of disregarding the $\mathfrak{sl}(2)$ subalgebra and accounting for the supercharge with one (negative) unit
\begin{equation}\label{RchargeCommutators}
    [L_0,J]=0,
    \quad
    [L_{\pm1},J]=0,
    \quad
    [J,G_{\pm1/2}]=-G_{\pm1/2}
\end{equation}
It can be motivated rigorously by looking at $\osp{2|1}$ as subalgebra of $\osp{2|2}$, where $J$ generates a $U(1)$ R-symmetry; see appendix \ref{appJ} for details. Due to $G_{\pm1/2}^2=L_{\pm1}$, the generator $J$ only has two distinct eigenvalues.

 Since $\mathfrak{sl}(2)$ forms a subalgebra of $\osp{2|1}$, we can obtain an intuition for the structure of the representations of $\mathfrak{osp}(1|2)$. A special feature encountered here, and generally arising for extended symmetries, is that various $\SL(2,\mathbb{R})$ representations are strung together into extended multiplets -- in our case superalgebra multiplets. Indeed, the fermionic supercharges $G_{\pm1/2}$ simply connect two $\mathfrak{sl}(2)$ lowest weight representations generated by $L_{-1}$, thereby providing a $\Z_2$-graded Hilbert space 
\begin{equation}\label{GradedKrylovSpace}
    \cH^{\osp{2|1}}_h=\cH_h^{\mathfrak{sl}(2)}\oplus\cH_{h+1/2}^{\mathfrak{sl}(2)}\,.
\end{equation}
The subspace $\cH_h^{\mathfrak{sl}(2)}$ is built on an $\mathfrak{osp}(2|1)$ lowest weight state $\ket{h}$ defined by
\begin{equation}
     L_{1}\ket{h}=0\,,
     \quad
     G_{1/2}\ket{h}=0\,,
     \quad
     L_0\ket{h}=h\ket{h}\,,
     \quad
     J\ket{h}=2h\ket{h}\,,
 \end{equation}
 and has the same Grassmann parity as $\ket{h}$.
 The subspace $\cH_{h+1/2}^{\mathfrak{sl}(2)}$ is reached by applying a supersymmetry transformation
\begin{equation}
    G_{-1/2}\ket{h}=\sqrt{2h}\ket{h+1/2}\,,
\end{equation}
where the normalised\footnote{The inner product for $\mathfrak{osp}(2|1)$ is inherited from its bosonic subalgebra $\mathfrak{sl}(2)$. For instance, $\bra{h}G_{1/2}G_{-1/2}\ket{h}=\bra{h}2L_0\ket{h}=2h$.} state $\ket{h+1/2}$ has Grassmann parity opposite to $\ket{h}$ and is the ground state of the $\mathfrak{sl}(2)$ highest weight representation $\cH_{h+1/2}^{\mathfrak{sl}(2)}$, as seen by $L_1G_{-1/2}\ket{h}=0$; it is no $\mathfrak{osp}(2|1)$ lowest weight state however. Observe that $\mathfrak{sl}(2)$ evolution as triggered by \eqref{Sl2dDisplacer} does not mix the subspaces $\cH^{\mathfrak{sl}(2)}_{h}$ and $\cH^{\mathfrak{sl}(2)}_{h+1/2}$, which is reserved for evolution by the supercharges $G_{\pm1/2}$. 

\begin{figure}[t]
    \centering
    \includegraphics[width=0.67 \linewidth]{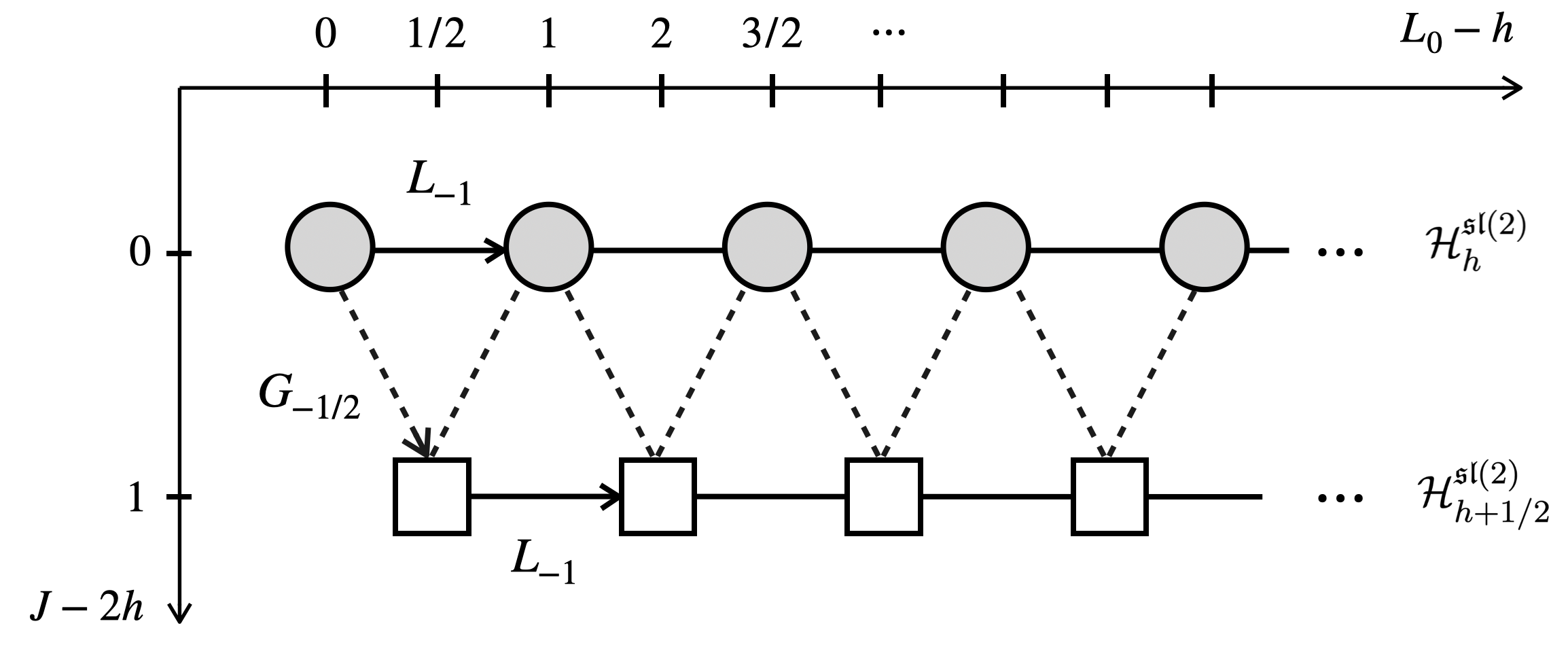}
    \caption{A highest weight representation of $OSp(1|2)$ couples two highest weight representations of $SL(2,\mathbb R)$ -- its bosonic subgroup -- by virtue of the (fermionic) supercharges. The horizontal direction indicates the bosonic sub-representations, whereas the vertical direction accounts for the fermion number. Note that the fermionic action is not nilpotent; rather $G_{\pm1/2}^2=L_{\pm1}$.}
    \label{fig:SUSY_hwrep}
\end{figure}

In this subsection, we study spread through the representation space \eqref{GradedKrylovSpace}. The quantum numbers of the commuting operators $(L_0,J)$ naturally organise this Hilbert space into a lattice depicted in figure \ref{fig:SUSY_hwrep}. Moreover, they serve naturally as position operators $X_i$. While $L_0$ counts progression in the horizontal direction of this figure, $J$ only detects spread in its vertical direction. We first study spread with a purely bosonic operator in subsection \ref{sec:bosOSP12Displacer} and, thereafter, by a fermionic operator in subsection \ref{OSPGrassmannOdd}. 

\subsubsection{Unitary \texorpdfstring{OSp$(2|1)$}{OSp12} supercoherent state}\label{sec:bosOSP12Displacer}

In this subsection, we study $OSp(1|2)$ displacement operators as found for instance in \cite{fatyga1989baker}. For our purposes, it suffices to settle on the particular unitary choice:
\begin{align}
D(\xi,\chi)
&=
\exp\left[\xi L_{-1}-\bxi L_{1}+\chi G_{-\frac{1}{2}}+\bchi G_{\frac{1}{2}}\right]\,,\label{ospDisplacer}
\end{align}
with Grassmann odd $\chi$ and $\bchi$, we are nevertheless able to describe the spread. In contrast to subsection \ref{sec:coh_superHW}, we start in this section with the dynamical algebra leading to the Krylov basis and Krylov operator, and discuss the lattice complexities afterward.

As above, we identify a dynamical spread algebra by a clever rewriting of the exponent in the displacement operator \eqref{ospDisplacer},
\begin{align}
    M_{1}
    =
    i&\frac{\bxi L_1-\bchi G_{1/2}}{\sqrt{|\xi|^2+\bchi\chi}}\,,
    \qquad 
    M_{-1}
    =
    -i\frac{\xi L_{-1}+\bchi G_{1/2}}{\sqrt{|\xi|^2+\bchi\chi}}\,,\\
    &M_0
    =
    L_0+\frac{\bxi \, \chi}{2|b|^2}G_{1/2}-\frac{\xi \, \bchi}{2|b|^2} G_{-1/2}\,,
\end{align}
with $\sl{2}$ commutators $[M_0,M_{\pm1}]=\mp M_{\pm}$, $[M_1,M_{-1}]=2M_0$. In their terms the displacement operator \eqref{ospDisplacer} assumes the form \eqref{Sl2dDisplacer} with $\alpha t\to\sqrt{|\xi|^2+\bchi\chi}$ and $\gamma\to0$. 

\begin{figure}[t]
    \centering
    \includegraphics[width=0.58\linewidth]{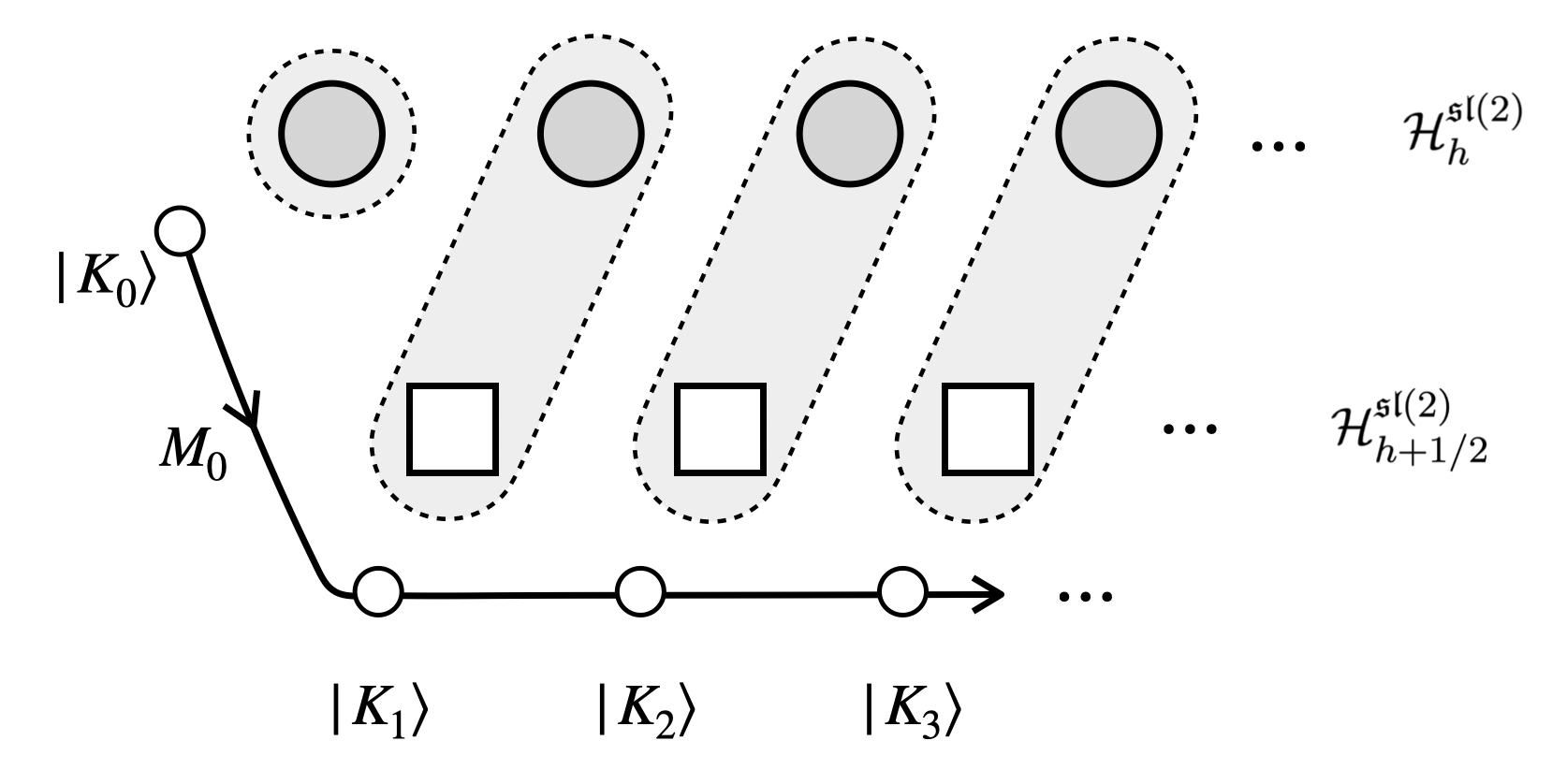}
    \caption{The $n^{\rm th}$ Krylov basis vector $\ket{K_n}$ combines the $n^{\rm th}$ state state in $\cH^{\sl{2}}_h$, drawn by circles, with the $n-1^{\rm st}$ state in $\cH^{\sl{2}}_{h+1/2}$, drawn by squares. The initial vector $\ket{K_0}$ has no partner. It is clear that $\cK\subset\cH^{\osp{2|1}}_h$. The specific linear combinations of circles and squares formed by $M_{-1}^n$ carve a \textit{Krylov path} through the weight lattice of $\cH^{\osp{2|1}}_h$.}
    \label{fig:dyn_sl2_osp12}
\end{figure}

Starting from the lowest weight state $\ket{h}$, the Krylov basis 
\begin{equation}\label{ospKbasis}
    \ket{K_n}
    =
    \frac{M_{-1}^n\ket{h}}{\norm{M_{-1}^n\ket{h}}},
    \qquad
    \norm{M_{-1}^n\ket{h}}^2
    =
    n!\frac{\Gamma(2h+n)}{\Gamma(2h)}\,,
\end{equation}
is constructed. It is easily seen by spelling out $M_{-1}^n$ in terms of $L_{-1}$ and $G_{-1/2}$ that $\ket{K_n}$ combines the $n^{\rm th}$ state in $\cH^{\sl{2}}_h$ with the $n-1^{\rm st}$ state in $\cH^{\sl{2}}_{h+1/2}$, as depicted in figure \ref{fig:dyn_sl2_osp12}. As with the super HW system studied above, these superpositions describe a Krylov path which the Krylov chain takes through the weight lattice \eqref{GradedKrylovSpace}. Observe that the Krylov space spans a strict subspace of the $\osp{2|1}$ representation, $\cK\subset \cH^{\osp{2|1}}$. 

The progression through the Krylov basis \eqref{ospKbasis} is given by the analogue of \eqref{Sl2Cplx}, where we require to additionally perform a semiclassical averaging as in \eqref{SpreadCplxSusy},
\begin{align}\label{CplxtyM0}
    \cC_{\osp{2|1}}^{(M_0)}
    &=
    \gev{\,\bra{\xi,\chi,h}M_0-h\ket{\xi,\chi,h}\,}\notag\\
    &=
    \gev{2h\sinh^2\left(\sqrt{|\xi|^2+\bchi\chi}\right)}
    =
    2h\sinh^2(|\xi(t)|)+h|\xi(t)|\sinh(2|\xi(t)|)|f(t)|^2\,,
\end{align}
From comparing the penultimate expression in \eqref{CplxtyM0} with \eqref{Sl2Cplx} we find the temporal dependence $\gev{|\xi|^2+\bchi\chi}=|\xi|^2+|f|^2=(\alpha t)^2$ for $\alpha\in\R$. Note that $\cC_{\osp{2|1}}^{(M_0)}\geq\cC_{\sl{2}}$ with equality only for $f=0$ or $\xi=0$. We see momentarily that the choice of $f$ is constrained.

\subsubsection*{Lattice complexities} 
Again, we investigate how the progression though the Krylov chain \eqref{ospKbasis} is measured \textit{with respect to the weight lattice} of the highest weight representation, i.e. as embedded into the total Hilbert space \eqref{GradedKrylovSpace}. As the lattice \eqref{GradedKrylovSpace} is labelled by $(L_0,J)$, these operators function naturally as position operators $\vec{X}$. As motivated below \eqref{SpreadCplxL}, we subtract their eigenvalue on the initial state $\ket{h}$, i.e. $\vec{X}=(L_0-h,J-2h)$. The computation of their lattice complexities \eqref{SpreadCplxL} is relegated to appendix \ref{app:spreadThroughWeightLattice} and only its results are reported here. The spread in the horizontal direction of figure \ref{fig:SUSY_hwrep} is
\begin{gather}
\begin{aligned}
   \cC_{\textrm{osp}(1|2)}^{(h)}(\xi,\chi) 
   &=\left|\int d\zeta_0 d\bzeta_0\, e^{\bzeta_0\zeta_0}\,\bra{\xi,\chi,h}L_0-h\ket{\xi,\chi,h}\right|
   \label{eq:OSP12cplx}\\
   &=
   \left|2h\sinh^2(|\xi|)
    +
    \frac{2h}{|\xi|^2}\bigl(\cosh(|\xi|)-\cosh(2|\xi|)+|\xi|\sinh(2|\xi|)\bigr)|f(t)|^2\right|\,.
\end{aligned}    
\end{gather}
where $\chi=f(t)\chi_0$ was chosen. By considering $\chi=0$ we clearly recover an $SL(2,\R)$ spread $\cC_{\textrm{SL}(2,\mathbb{R})}^{(h)}(\xi)=2h\sinh^2(|\xi|)$. Observe that the last term is proportional to the additional term in \eqref{CplxtyM0}. However, the overall structure of the additional terms is different. Using that $\chi=f(t)\chi_0$, the spread in the vertical direction of the lattice in figure \ref{fig:SUSY_hwrep} is,
\begin{align}\label{Jspread}
    \cC_{\osp{2|1}}^{J}
    =\left|\int d\zeta_0 d\bzeta_0\, e^{\bzeta_0\zeta_0}\,
    \bra{\xi,\chi,h}J-2h\ket{\xi,\chi,h}\right|
    =
    P^{(h+1/2)}
    =
    4h\frac{\cosh(|\xi|)-1}{|\xi|^2}|f(t)|^2\,.
\end{align}
$P^{(h+1/2)}$ is defined in \eqref{ProbSectors} and is the probability of residing in the sector $\cH^{\mathfrak{sl}(2)}_{h+1/2}$. Since this takes a value between zero and one, we find a necessary constraint on the choice of $f$,
\begin{equation}
    0\leq|f|^2\leq \frac{|\xi|^2}{4h(\cosh(|\xi|-1))}
\end{equation}

\subsubsection*{Remark: The purely Grassmann case}
It is interesting to consider the evolution for a pure supercharge transformation. To do so we simply have to set $\xi=0$ in the displacement operator in eq. \eqref{ospDisplacer}. This has the vacuum state evolve to
\begin{equation}
    \ket{h}\to \ket{\chi,h}:=\ket{0,\chi,h}
    =
    e^{\chi G_{-1/2}+\bchi G_{1/2}}\ket{h}
    =
    (1-h\bchi\chi)\ket{h}+\chi G_{-1/2}\ket{h}\,.
\end{equation}
This evolution truncates after hitting only one state besides the vacuum state. Of course this happens due to the Grassmann oddness of $\chi$, since $G_{\pm1/2}$ are not nilpotent. Its complexities are computed easily 
\begin{align}
   \cC_{\textrm{osp}(1|2)}^{(h)}(0,f(t)\zeta_0) 
   &=
   \int d\zeta_0 d\bzeta_0\, e^{\bzeta_0\zeta_0}\,\bra{\chi,h}L_0-h\ket{\chi,h}
   =
   h|f(t)|^2\,,\\
    \cC_{\textrm{osp}(1|2)}^{(h)}(0,f(t)\zeta_0) 
   &=
   \left|\int d\zeta_0 d\bzeta_0\, e^{\bzeta_0\zeta_0}\,\bra{\chi,h}J-2h\ket{\chi,h}\right|
   =
   2h|f(t)|^2\,.
\end{align}
This is now evidently similar to the simpler construction in subsection \ref{secGrassmannOddParameter}. The reason is clear: the fact that $\chi$ is Grassmann odd stifles the evolution after only two Krylov steps. It is therefore interesting to study a case where the supercharges are accompanied by a Grassmann even parameter. This is the subject of the next subsection. We find that the changes are drastic.

\subsubsection{Supercharge induced spread}\label{OSPGrassmannOdd}
The ingredients setting apart the $\mathfrak{osp}(2|1)$ algebra from its bosonic subalgebra $\sl{2}$ are the supercharges $G_{\pm1/2}$. It is thus interesting to study the spread they induce through the Hilbert space $\cH^{\osp{2|1}}_h$ in \eqref{GradedKrylovSpace}. To study the evolution driven by a supercharge in isolation, we pick the Hermitian combination
\begin{equation}\label{SusyLiouvillian}
    \cL=\alpha(G_{-1/2}+G_{1/2})\,,
\end{equation}
with a Grassmann even parameter $\alpha\in\R$.  As a result, this operator is total Grassmann odd object now, and therefore its exponential $e^\cL$ is not an element of the supergroup $OSp(2|1)$ -- in contrast to the situation in subsection \ref{sec:bosOSP12Displacer}. The coherent state approach used in this paper thus not straightforwardly applicable. In this subsection, we show that this evolution can nevertheless be studied by use of the Lanczos algorithm. As with the simple case studied on page \pageref{secGrassmannEvenParameter} the Krylov basis alternates between Grassmann even and odd vectors. 

\begin{figure}
    \centering
    \includegraphics[width=0.4\linewidth]{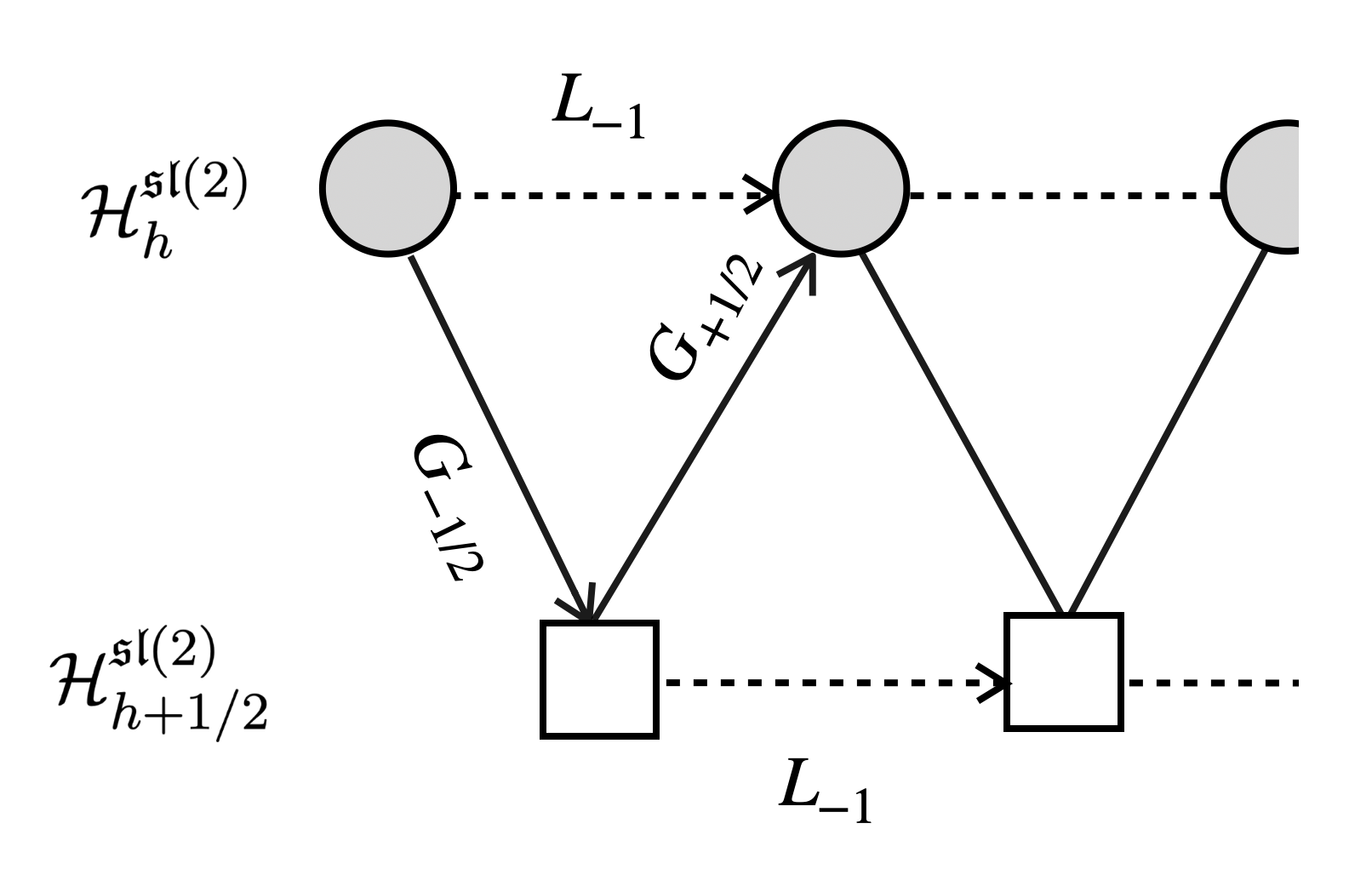}
    \caption{The operator evolves the system in a `zig-zag' motion, following the solide line, alternating between Grassmann even and odd states following the solid lines, which indicate the Krylov path. In this case the Krylov space exhausts the entire weight lattice, $\cK=\cH^{\osp{2|1}}_h$. }
    \label{fig:chain_squeezer_fermOsp}
\end{figure}

The Lanczos algorithm is reviewed in appendix \ref{app:Lanczos_survival} and performed for initial state $\ket{K_0}=\ket{h}$ and the operator \eqref{SusyLiouvillian} in appendix \ref{appSusyKrylov}. Denoting the norm of a vector in Hilbert space by $\norm{\ket{v}}=\sqrt{\braket{v}}$, the emerging pattern for the Krylov basis is
\begin{equation}\label{SusyKrylov2}
    \ket{K_n}=\frac{G_{-1/2}^n\ket{h}}{\norm{G_{-1/2}^n\ket{h}}}\,,
    \qquad
    b_n=\alpha\frac{\norm{G_{-1/2}^n\ket{h}}}{\norm{G_{-1/2}^{n-1}\ket{h}}}\,.
\end{equation}
Evidently these states are all orthogonal since they lie in distinct energy eigenspaces, each shifted from their neighbour in half-integral steps. The Lanczos coefficients $b_n$ trade the normalisation of the preceding Krylov basis element $\ket{K_{n-1}}$ for that of $\ket{K_n}$. In contrast to the evolution studied in the previous subsection, the Krylov space generated from \eqref{SusyLiouvillian} exhausts the entire highest weight representation \eqref{GradedKrylovSpace}, that is, $\cK=\cH^{\osp{2|1}}_h$, and the Krylov path is depicted in figure \ref{fig:chain_squeezer_fermOsp}.

Indeed, for even $n$, the state $\ket{K_n}$ shares the Grassmann parity of $\ket{h}$ and belongs to the even graded part of the Hilbert space $\cH^{\mathfrak{sl}(2)}_{{h}}$, while for odd $n$ it has opposite Grassman parity and lies in $\cH^{\mathfrak{sl}(2)}_{h+1/2}$. The Lanczos coefficients in \eqref{SusyKrylov2} also mirror the $\Z_2$ gradation of \eqref{GradedKrylovSpace}. As derived in appendix \ref{appSusyKrylov}, they differ for even $n=2k$ and odd $n=2k+1$,
\begin{equation}\label{OSP12Lanczos}
    b_{2k}
    =
    \alpha\frac{\norm{G_{-1/2}^{2k}\ket{h}}}{\norm{G_{-1/2}^{2k-1}\ket{h}}}
    =
    \alpha\sqrt{k}\,,
    \qquad
    b_{2k+1}
    =
    \alpha^2\frac{\norm{G_{-1/2}^{2k+1}\ket{h}}}{\norm{G_{-1/2}^{2k}\ket{h}}}
    =
    \alpha\sqrt{2h+k}\,.
\end{equation}
Such Lanczos coefficients have been encountered before in \cite{muck2022krylov}, where the Krylov complexity associated with Laguerre polynomials was evaluated. While their analysis is purely mathematical, our current discussion presents an explicit realisation of their analysis. Physical models to which the present discussion can be applied are easily found: For instance in the tri-critical Ising model, which features $\cN=1$ superconformal symmetry. Adapting the parameters of the complexity presented \cite{muck2022krylov} for the coefficients in \eqref{OSP12Lanczos}, we can directly quote the complexity,
\begin{align}
    \mathcal{C}_S(t)
    %
    &=
    \frac{(\alpha t)^2}{2}+\frac{4h-1}{4(2h-1)}\left[1-e^{-(\alpha t)^2}\,{}_2F_2\left(\frac{3}{2}-2h;\frac{1}{2};(\alpha t)^2\right)\right]\label{SusyCplx}\\
    &\overset{t\to\infty}{=}
       \frac{(\alpha t)^2}{2}
       +
       \frac{4h-1}{4(2h-1)}\left[1-\frac{\sqrt{\pi}}{\Gamma(3/2-2h)}(\alpha t)^{2-4h}\right]\,,\label{SusyCplxLateTime}  
\end{align}
where ${}_2F_2$ is a generalised hypergeometric function. The second line exhibits the late-time behaviour of the complexity. 

As already explained in \cite{muck2022krylov}, the complexity oscillates at early times about $t^2$, and asymptotes toward this behaviour at late times. Recalling the purely fermionic HW coherent state in section \ref{sec:FermionicHW}, where the Grassmann even parameter gave rise to an oscillatory spread, whereas a Grassmann odd parameter gave rise to a spreading proportional to $t^2$, we find that the spread complexity \eqref{SusyCplx} describes a hybrid thereof.

Before closing this section, note that upon squaring the operator in eq. \eqref{SusyLiouvillian} we find
\begin{equation}\label{LSquare}
    \cL^2=\alpha^2(L_{-1}+L_1)+2\alpha^2L_0\,,
\end{equation}
where \eqref{ospAlgebra} has been employed. This is purely bosonic and of the form used in the $\mathfrak{sl}(2)$ displacement operator \eqref{Sl2dDisplacer}. For the complexity \eqref{Sl2Cplx}, this situation presents a singular limit, where exponential growth of \eqref{Sl2Cplx} truncates to quadratic growth. Having $\cL^2$ evolve the ground states of $\cH^{\mathfrak{sl}(2)}_h$ and $\cH^{\mathfrak{sl}(2)}_{h+1/2}$, respectively, leads to a spread complexity
\begin{equation}
    \mathcal{C}^{\cL^2}_{h}(t)
    =
    2h\alpha^4t^2\,,
    \qquad
    \mathcal{C}^{\cL^2}_{h+1/2}(t)
    =
    2(h+1/2)\alpha^4t^2\,.
\end{equation}
Hence, these complexities have the same late time scaling as the complexity given in \eqref{SusyCplxLateTime}\footnote{We restrict here to unitary representations for which $h>0$. For non-unitary representations, we see that the scaling can be larger in \eqref{SusyCplxLateTime}.}.

\section{Spread complexity for semiclassical strings in \texorpdfstring{$AdS_5\times S^5$}{AdS5xS5}}\label{sec:semicl_spin_chains}

We now turn to the main result of this paper: computing the spread complexity for semiclassical string states, and their holographic duals, in the planar limit of the $AdS_5$/CFT${}_4$-correspondence. 

In this section we briefly remind the reader of some crucial ingredients mapping coherent states on the gauge and string theory sides. We then go on to compute spread complexity of so-called rotating strings in different parts of the $AdS_5\times S^5$-geometry\footnote{Note that similar spin chains have studied by \cite{Rabinovici:2022beu}, though in a different context. In that paper the authors discusses \textit{operator} Krylov complexity in Heisenberg and related spin chains, here we consider spread complexity in the continuum limit of the \textit{semiclassical} regime of such spin chains, eventually leading the their corresponding string states.}.

In each case, by determining the trajectory in phase space through the equations arising from the continuum limit of the pertinent spin chain Hamiltonian, we identify the time and space dependence of the coherent state. With this in hand, we compute the return amplitude and subsequently extract the Lanczos coefficients, as reviewed in Appendix \ref{app:Lanczos_survival}. Here, the pattern observed in section \ref{sec:superC} emerges once again: spread complexity manifests itself effectively as a one-dimensional Krylov path.

\subsection{The coherent states  -- large charge string states correspondence}

In the planar limit of the AdS/CFT correspondence, the operator of interest on the gauge theory side are single trace operators built from scalars in $\cN=4$ SYM. The total symmetry group of $\mathcal N= 4$ SYM theory combines the conformal symmetry, the supersymmetery and the R-symmetry into the superconformal group $PSU(2,2|4)$. 
This theory admits six scalars $\Phi^a$, $a=1,\dots, 6$, which are often packaged into three complex scalars $X=\Phi^1+i\Phi^2$, $Y=\Phi^3+i\Phi^4$ and $Z=\Phi^5+i\Phi^6$. Restricting for now to two complex scalars for simplicity, the generic single trace operator takes the form
\begin{align}\label{eq:operator_example}
    \mathcal O_{J_1,J_2}=\mathrm{Tr}(ZZXZXX\cdots ZX)\,,
\end{align}
where $J_1$ and $J_2$ labels the numbers of appearances in the trace of $X$'s and $Z$'s, respectively. The free theory conformal dimension of this operators is $\Delta_0=J_1+J_2$, while the 1-loop anomalous dimension is captured by the (1-loop) dilatation operator\footnote{Although in general, interactions in the gauge theory lead to mixing with non-scalar operators. at one loop mixing this does not occur \cite{Minahan:2002ve} and one can restricted to operators within a certain ``closed'' sector.}. More generally, starting from $\mathcal N=4$ SYM theory, one can consider single trace operators associated to the $\mathcal N=1$ superfields which take on the schematic form
\begin{align}\label{eq:operator_su23}
    \mathcal O=\mathrm{tr}(X^{J_X}Y^{J_Y}Z^{J_Z}\psi_1^{K_1}\psi_2^{K_2})\,,
\end{align}
where besides the three scalars of the matter supermultiplets one also includes two spinor components  $\psi_i$ of the gaugino supermultiplet. These form an $SU(2|3)$-subsector of the SYM theory \cite{Beisert:2003ys,Staudacher:2004tk}.

In \cite{Minahan:2002ve}, the one-loop dilatation operator was famously identified as an integrable spin chain Hamiltonian, reducing the computation of the  1-loop anomalous dimensions in the scalar sector to that of diagonalising the spin chain Hamiltonian using the Bethe Ansatz.  Focussing first on the subsector $\mathfrak{su}(2)\subset \mathfrak{psu}(2,2|4)$, the anomalous dimensions at one-loop are determined by diagonalising the operator
\begin{align}
    \mathcal D_{\mathfrak{su}(2)}=\frac{\lambda}{8\pi^2}\sum_{k=1}^L(1-P_{k,k+1})\,,
\end{align}
where $\lambda$ is 't Hooft coupling.
It is well-known that this is the Hamiltonian for a periodic spin chain with $L$ sites, as evident after rewriting it in terms of spin operators
\begin{align}
    H_\mathrm{XXX}\equiv\mathcal D_{\mathfrak{su}(2)}=\frac{\lambda}{8\pi^2}\sum_{k=1}^L\left(\frac{1}{2}-2\vec \sigma_k\cdot \vec \sigma_{k+1}\right)\,,
\end{align}
where $\vec \sigma_k$ is a vector of the three Pauli matrices $\sigma^i$ for each site $k$. This is the Hamiltonian of the well-known XXX-spin chain. In this picture, the single trace operator \eqref{eq:operator_example} becomes an excited state above the ferromagnetic ground state. For example identifying $X$ as spin up and $Z$ as spin down, the $Z$-inserting are excitations/impurities onto the ground state of all spin-up states. It is insightful for later to re-express this Hamiltonian in terms of the raising and lowering generators
\begin{align} \label{eq:spin_chain_H_su2_rl}
     H_\mathrm{XXX}= \frac{\lambda}{8\pi^2}\sum_{k=1}^L\left[2\Big((J_+)_k(J_-)_{k+1}+(J_-)_k(J_+)_{k+1}\Big)+(J_0)_k(J_0)_{k}\right]\,.
\end{align}
where $\sigma_\pm =(\sigma^x\pm i\sigma^y)$ and $\sigma^z=J_0$.
The general dilatation operator in the $\mathfrak{su}(m|n)$ sector is 
\begin{align}\label{eq:spin_chain_H_sunm}
    \mathcal D_{\mathfrak{su}(m|n)}=\frac{\lambda}{8\pi^2}\sum_{k=1}^L\left(\frac{m-n-1}{m-n}-\sum_{A,B}g_{AB}(T_A)_k(T_B)_{k+1}\right)\,,
\end{align}
where $T_A$ are the generators of the superalgebra $\mathfrak{su}(m|n)$, the subscript indicates they are attached to a fixed site on the chain and $g_{AB}=\mathrm{Str}(T_AT_B)$ is the Killing-metric on the Lie superalgebra.

In \cite{Kruczenski:2003gt}, the author showed that the spin chain Hamiltonian plays an even deeper role in the AdS/CFT correspondence. Taking an appropriate semiclassical limit, the spin state of the spin chain can be described by coherent states, one at each site of the chain. Taking a continuum limit of the spin chain Hamiltonian in the coherent state representation, these can be identified with semiclassical fast spinning string states on the gravity side. In this semiclassical regime the state at site $k$ of the spin chain is described by
\begin{align}\label{eq:coherent_state_single_site}
    |\vec z_k\rangle =\mathcal N_k \exp(D_k(\vec z_k))|\Lambda_k\rangle\,,
\end{align}
where $\mathcal N_k $ is the normalisation factor and $D_k(\vec z_k)$ is the displacement operator constructing the coherent state for the Lie algebra symmetry $\mathfrak g$ of the spin chain under consideration.
That is, on the gauge side, we can use coherent states to model the spin states at each site, such that the total state for the whole spin chain is
\begin{align}\label{eq:spinchainstate}
    |\vec z\rangle \equiv \bigotimes_{k=1}^L |\vec z_k\rangle\,,
\end{align}
where $\vec z_k$ is a vector of phase space parameters of the $k$-coherent state given in  eq. \eqref{eq:coherent_state_single_site}.

Following the steps reviewed in section \ref{sec:semiclassics}, 
the path integral approach leads to a semiclassical action of the form given in eq. \eqref{eq:semi_class_act}.  Subsequently, taking the long wave length limit, the discrete number of sites become a continuum space coordinate $\sigma$. The action then becomes a  (Landau Lifshitz) sigma model, up to corrections $1/L$
where $L$ is the length of the chain. In this continuum limit, which corresponds holographically to the BMN  limit, the SYM operators with large charge can be identified with the propagation of a closed string. In this identification,  spin chain coherent excitations precisely capture the string profile where the symmetry group taken as a target space, and vice-versa.

\subsection{Generalising spread complexity to semiclassical spin chains}\label{sec:gen_spread_complex_spinchains}

\noindent
Before moving on to the computation of spread complexities, let us pause to stress the new and distinct features of spread complexity when applied to this semiclassical sector of the holographic correspondence.

\begin{figure}[t]
    \centering
    \includegraphics[width=0.62\linewidth]{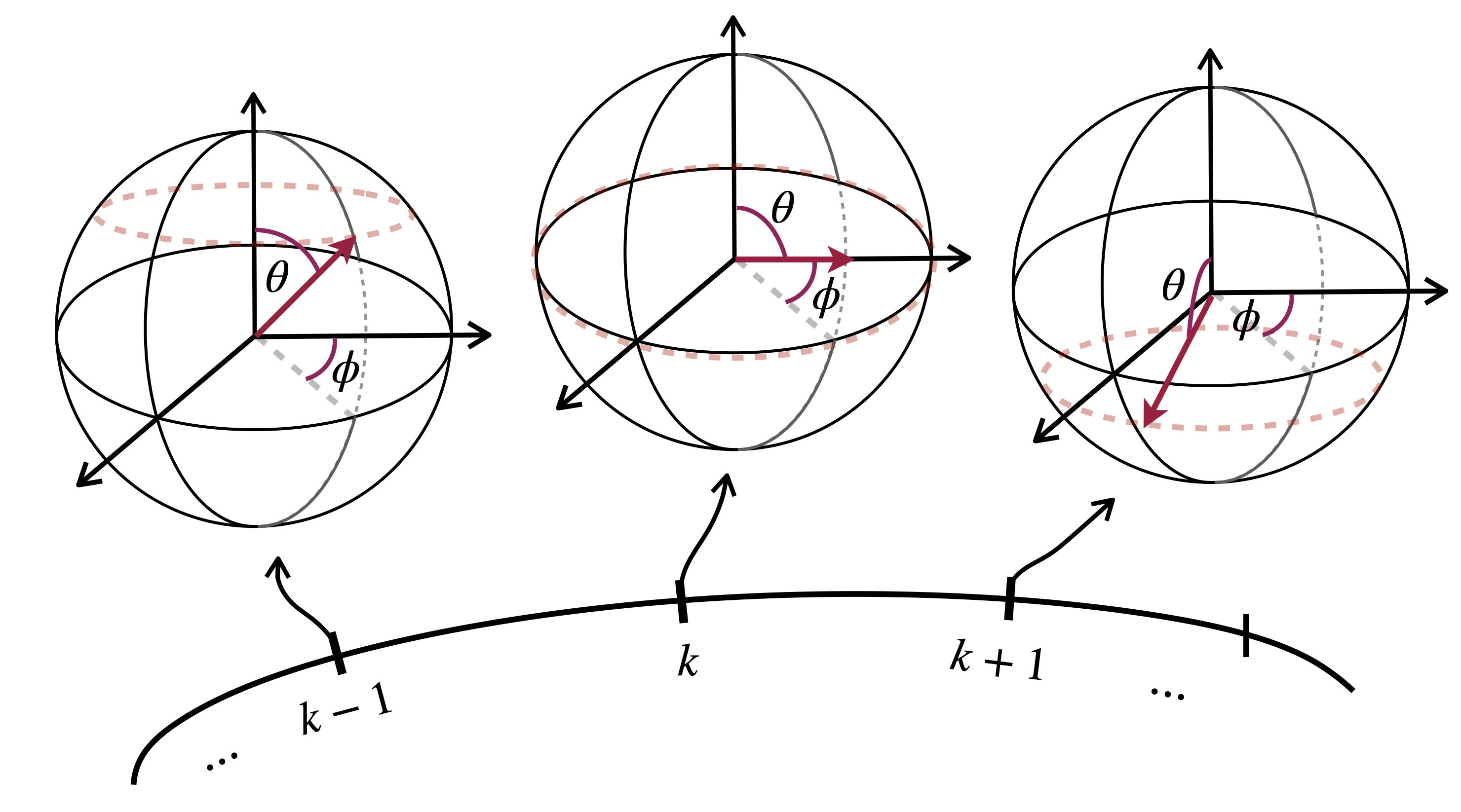}
    \caption{An approximate illustration of semiclassical spin chain. In the semiclassical limit, the spins at each single site of the lattice are approximated by a coherent state. Taking the continuum limit, the resulting equations of motion lead to a one-dimensional trajectory in the coherent state's the phase space. For simplicity, we only depict the coherent states at the original spin chain site, but one should imagine a continuum of coherent states along the length of the spin chain.}
    \label{fig:coh_spc}
\end{figure}

\begin{itemize}
    \item   In contrast to section \ref{sec:fermion_osc} and \ref{sec:superC}, the initial state (on each site of the spin chain) from which we compute spread complexity is no longer the vacuum state or equivalently the lowest weight state. Instead, the initial state will be a coherent state -- one for each site on the chain. Since each site on the chain carries an individual coherent state, the initial state is a tensor product of $L$ independent coherent states as given in eq. \eqref{eq:spinchainstate}.

    \item In addition, we no longer evolve the initial state using the exponent of the displacement operators. The semiclassical evolution is dictated by the spin chain Hamiltonian, with symmetry corresponding to the coherent state and vice-versa. Since the Hamiltonian is quadratic in the generators, the resulting semiclassical equations are now a set of \textit{second} order differential equations. 
    Inspecting the spin chain Hamiltonian in eq. \eqref{eq:spin_chain_H_su2_rl}, we see that it acts \textit{linearly} in the generator on the $k^{\mathrm{th}}$ and $(k+1)^{\mathrm{th}}$ individually.  The critical observation is that here though the action in each entry are coupled: raising in one entry while lowering in its neighbour, and vice-versa. 

    \item Finally, instead of a single state, we consider a set of $L$ states evolving with nearest-neighbour interactions, which, in the continuum limit, introduces additional space dependence. In particular, the resulting complexity does generically not only carry a time-dependence but also a non-trivial space-dependence. The situation is sketched in figure \ref{fig:coh_spc}. In fact, we shall also encounter spread complexities which are constant in time, but admit a non-trivial spatial profile. 
\end{itemize}
All these new elements lead to a much richer set of solutions, leading to multiple possible trajectories in phase space with a larger set of semiclassical trajectories for each individual spin chain site, see fig. \ref{fig:hyperbolic_disc}.

\subsection{Spread complexity of the rotating string}
\label{subsec:sc_rs}

A set of key solutions in the semiclassical regime of the planar sector of holography are so-called rotating strings. These are nearly point-like strings spinning at large momentum in a submanifold of $AdS_5\times S^5$, and thus corresponding to a different subgroup of the supergroup $PSU(2,2|4)$. In this section, we compute spread complexity via the pertinent survival amplitude\footnote{In appendix \ref{app:Lanczos_survival}, we review how the Lanczos coefficients and the spread complexity are obtained from the moments.} for different (super) string states or, equivalently, for a set of long operators, i.e. captured by spin chains of different closed subsectors of the symmetry group $\mathfrak{psu}(2,2|4)$ of the $AdS_5\times S^5$-correspondence.

Remarkably, despite the differences summarised in section \ref{sec:gen_spread_complex_spinchains} with respect to cases studied so-far in the literature and above, spread complexity materialises itself again in terms of an effective $SL(2)$ or $SU(2)$-algebra. Indeed, we consistently extract the Lanczos coefficients for rotating strings in, respectively, the compact and non-compact subsectors of the $\frak{psu}(2,2|4)$-symmetry algebra. The important observation is that this result is far from implying that the associated spread complexity trivially reduces to that $SL(2)$ or $SU(2)$-coherent states of \cite{Caputa:2021sib}. Since the dynamics is driven by the corresponding spin chain Hamiltonian, the time-dependence of the complexity is captured by an entirely new trajectory in the phase space of the coherent state. This in particular, leads to complexities which take the functional form of an $\sl{2}$ or $\su2$ complexity -- we refer to this as an \quotes{atomic pattern} --  but different time-evolutions than spread complexities obtained in \cite{Caputa:2021sib}. 

Turning finally, to a fermionic sector, we arrive at a similar conclusion: spread complexity again follows an atomic pattern, now one that closely follows the spread complexity of multi-fermion states uncovered in section \ref{sec:fermion_osc}. In addition, for all the considered cases, the emergent spatial dependence, inherited from taking the continuum limit of the spin chain sites, leads to a spatial profile to the time-dependent spread complexity.

\subsubsection*{$SU(2)$-subsector}
The simplest closed subsector consists out of the $SU(2)$ XXX or Heisenberg spin chain. The corresponding strings are positioned in the centre of $AdS_5$ while rotating on $S^5$ within two out of three planes, that is they have angular momenta  $S_1,S_2 = 0\,,J_3 =
0\,, J_1,J_2 \neq 0$ in the notation of \cite{Stefanski:2004cw}.
For completeness let us state the $SU(2)$-coherent state 
\begin{align}
    |z,j\rangle&=(1+|z|^2)^{-j}\sum_{n=0}^{2j} z^n\sqrt{\frac{\Gamma(2j+1)}{n!\Gamma(2j-n+1)}}|j,-j+n\rangle\\
    &=\frac{1}{\cos(\theta/2)^{2j}}\sum_{n=0}^{2j}e^{in\phi}\tan(\theta/2)^n\sqrt{\frac{\Gamma(2j+1)}{n!\Gamma(2j-n+1)}}|j,-j+n\rangle\,.
\end{align}
The total spin chain coherent state is of course the $L$-tensor product $|\{(z_k,j_k)\}_{k=1}^L\rangle=\bigotimes_{k=1}^L|z_k,j_k\rangle$. In order to identify a valid trajectory in phase space, we turn to the continuum equations of motion. They take the form \cite{Kruczenski:2003gt}
\begin{gather}
\begin{aligned}\label{eq:eqs_cont_su2}
    \sin(\theta)\partial_t \theta+\lambda j \partial_\sigma (\sin^2\theta\partial_\sigma \phi)&=0\,,\\ 
    \sin(\theta)\partial_t \phi+\lambda j \partial_\sigma^2\theta -\lambda j \sin\theta\cos\theta(\partial_\sigma \phi)^2&=0\,,
\end{aligned}
\end{gather}
and are additionally subjected to the periodic boundary conditions $\phi(\sigma=J,t)=\phi(\sigma=0,t)$ and $\theta(\sigma=J,t)=\theta(\sigma=0,t)$.

Using the Ansatz $\partial_\sigma\phi=0$, the author of \cite{Kruczenski:2003gt} found rotating string solutions satisfying the zero momentum constraint imposed by the cyclicity of the trace in \eqref{eq:operator_example}. From the equations in \eqref{eq:eqs_cont_su2}, one obtains that $\partial_t\theta=0$ and $\partial_t^2\phi=\omega$. From the latter one assumes in addition that $\partial_t\phi=\omega$, the angle dependence on $\sigma$ is captured by the equation
\begin{align}
    \partial_\sigma^2 \theta=-\frac{\omega}{\lambda j}\sin \theta\,.
\end{align}
This integrates to $\partial_\sigma \theta = \pm \sqrt{a + b \cos \theta}$, where $a$ is an integration constant and $b = 2\omega/{\lambda}$. One can distinguish two cases \cite{Kruczenski:2003gt}, leading to circular and folded rotating strings respectively. Either  $a > |b|$, and the angle $\theta$ is unconstrained or when \(b > |a|\), implying that for the value $\theta_0 = \arccos(-\frac{a}{b})$ the square root vanishes. The latter leads to a point-like string state oscillating between $-\theta_0 < \theta < \theta_0$.

We can choose $\phi=\omega t$ and $\theta_i=\theta(\sigma)$ with a $\sigma$-dependent profile. The corresponding coherent state for the $j=1/2$-representation is 
\begin{align}
    |(z,1/2), t\rangle &=\frac{1}{\cos(\theta(\sigma)/2)}\sum_{n=0}^{1} e^{in\omega t}\tan(\theta(\sigma)/2)^n|1/2,-1/2+n\rangle\,.\label{eq:SU2_rotating}
\end{align}
The state of the spin chain with $L$ number of sites at time $t$ is a tensor product of $L$ copies of \eqref{eq:SU2_rotating}. To calculate the complexity of the time evolution associated with the state $\bigotimes_{k=1}^L |z_k,1/2,t\rangle$, we compute the autocorrelation function, $S(t)= \langle \{(z_k,j_k)\}_{k=1}^L,  t | \{(z_k,j_k)\}_{k=1}^L, 0\rangle$ and obtain the following Lanczos coefficients by the moment recursion method,
\begin{equation}
    a_n=\omega(1-2\cos ^2(\theta(\sigma)/2))(n-L/2)-\frac{m\omega}{2},~~b_n=\frac{\omega \sin{(\theta(\sigma))}}{2}\sqrt{n(n+L-1)} \label{eq:SU_RS_LC}
\end{equation}
This set of Lanczos coefficients can be mapped to the set of Lanczos coefficients of \textit{single} $SU(2)$ coherent state by the following parametrisation, $\alpha=\frac{\omega \sin{(\theta_i(\sigma))}}{2}$, $h=\frac{L}{2}$, $\gamma=\omega(1-2\cos ^2(\theta_i(\sigma)/2))$ and $\delta=-\frac{m\omega}{2}$. This allows us to recast the Lanczos coefficients in the form, 
\begin{equation}
    a_n = \gamma(-j_{\mathfrak{su}(2)} +n)+\delta, \quad b_n = \alpha\sqrt{n(-n+2j_{\mathfrak{su}(2)} +1)}
\end{equation}
and obtain the corresponding spread complexity as,
\begin{eqnarray}
    \mathcal{C}(t,\sigma)= \frac{2j}{1+\frac{\gamma^2}{4\alpha^2}} \sin^2\left(\alpha t \sqrt{1+\frac{\gamma^2}{4\alpha^2}}\right)=\frac{L}{\csc ^2(\theta(\sigma) )}\sin{\left(\alpha t\csc (\theta(\sigma) )\right)}^2,
\end{eqnarray}
where $\sqrt{1+\frac{\gamma^2}{4\alpha}}=\csc (\theta_i(\sigma) )$ for the Lanczos coefficients of \eqref{eq:SU_RS_LC}. This expression, characterising the spread complexity of a string state, carries now a worldsheet dependence. In particular, the spread complexity is now a time-dependent function, measuring the spread in the corresponding Hilbert space, with an additional spatial profile. The resulting complexity is oscillatory with respect to time where the period of oscillation is governed by the choice of the initial value of the $\theta_i$ parameter. This comes as no surprise since the number of states on each sites is finite. To be in the continuum limit, we take the length of the chain $L$ to be sufficiently large but finite. This expression then quantifies the spread complexity for both a string state propagating in $S^3\subset S^5$ as well as that for single trace operators as in eq. \eqref{eq:operator_example}, up to $1/L$-corrections.

\subsubsection*{$SU(3)$ subsector}
The next simplest case, is to consider the states for the subalgebra $\mathfrak{su}(3)$ of $\mathfrak{psu}(2,2|4)$ \cite{Stefanski:2004cw,Hernandez:2004uw}. This sector corresponds on the gauge side to single trace operator with all three scalars included.
Let us immediately restrict to the relevant case, i.e. $j=1/2$. Then the $SU(3)$-coherent state reads
\begin{align}
|n\rangle = \cos \theta e^{i\varphi} |1\rangle + \cos \psi \sin \theta e^{-i\varphi} |2\rangle + \sin \psi \sin \theta e^{i\phi} |3\rangle\,,
\end{align}
Rotating string solutions correspond to the ansatz \cite{Hernandez:2004uw}
\begin{align}
    \theta = \theta_0 \,,\quad \psi = \psi_0\,,\quad    \partial_\sigma \varphi = m \,,\quad  \partial_\sigma \phi = n\,,
\end{align}
where $m$ and $n$ are integers, this solves the equations of motion with angular momentum 
\begin{align}
   \dot{\varphi} &= -\frac{\lambda}{J^2} \left[ m \left( \frac{J_1}{J} - \frac{J_2}{J} \right) + n \frac{J_3}{J} \right]\equiv\omega_\phi\,,\\
   \dot{\phi} &= \frac{\lambda}{2J^2} \left[ (n^2 - m^2 - 2n) \left( \frac{J_1}{J} - \frac{J_2}{J} \right) + n \frac{J_3}{J} \right]\equiv \omega_\phi\,,
\end{align}
where the different constants are related to the fixed angles $\theta_0$ and $\psi_0$ via
\begin{align}
    \frac{J_1 - J_2}{J_1 + J_2} = \cos(2\psi_0)\,,\qquad  \frac{J_3}{J} = \sin^2 \theta_0\,.
\end{align}
In other words the coherent state is 
\begin{align}
|n\rangle = \cos \theta_0 e^{i(m\sigma +\omega_\varphi t)} |1\rangle + \cos \psi_0 \sin \theta_0 e^{-i(m\sigma +\omega_\varphi t)} |2\rangle + \sin \psi_0 \sin \theta_0 e^{i(n\sigma +\omega_\phi t)} |3\rangle\,,\label{eq:su3_rs}
\end{align}
To quantify the complexity associated with the time evolution of the state \eqref{eq:su3_rs}, we compute the autocorrelation function $S(t)$ and derive the corresponding Lanczos coefficients using the moment recursion method as follows:
\begin{equation}
 a_n = (n-L/2) 2\omega f(\theta,\psi), \quad b_n =  \sqrt{n(L-n+1) \omega ^2 g(\theta,\psi)}\label{eq:SL_RS_LC2}  
\end{equation}
where the functions $f(\theta,\psi)$ and $g(\theta,\psi)$ are respectively given by $\left(\cos ^2(\theta )-\sin ^2(\theta ) \cos (2 \psi )\right)$ and $ \sin ^2(\theta ) \cos ^2(\psi ) \left(-2 \sin ^2(\theta ) \cos (2 \psi )+\cos (2 \theta )+3\right)$. Note that, the return amplitude $S(t)=\langle n(t)|n(0)\rangle$ is independent of $\sigma$, and so is the corresponding complexity.  These Lanczos coefficients can be mapped to those of the $SU(2)$ coherent state using the parametrisation $\alpha = \omega\sqrt{g(\theta,\psi)}$, $j_{\mathfrak{su}(2)} = L/2$, $\gamma = 2\omega f(\theta,\psi)$, and $\delta = 0$. Under this parametrisation, the Lanczos coefficients become
\begin{equation}
    a_n = \gamma(-j_{\mathfrak{su}(2)} +n), \quad b_n = \alpha\sqrt{n(-n+2j_{\mathfrak{su}(2)} +1)}.
\end{equation}
The corresponding spread complexity is expressed as
\begin{eqnarray}
    \mathcal{C}(t) = \frac{2j}{1+\frac{\gamma^2}{4\alpha^2}} \sin^2\left(\alpha t \sqrt{1+\frac{\gamma^2}{4\alpha^2}}\right) = \frac{L}{1+\frac{f(\theta,\psi)^2}{g(\theta,\psi)}}\sin^2\left(\alpha t \, \sqrt{1+\frac{f(\theta,\psi)^2}{g(\theta,\psi)}}\right).
\end{eqnarray}
The resulting complexity again exhibits oscillatory behaviour in time, with the oscillation period governed by the ratio $f(\theta,\psi)^2/g(\theta,\psi)$. Here, $L$, the length of the spin chain plays the role of the total spin of the system and determines the dimension of the Krylov space. Note that once again the Krylov space is finite, truncating at a dimension of $L+1$.

\subsubsection*{$SU(1,1)$ or $SL(2)$-subsector}
The $SL(2)$-subsector leads again to a dilatation operator that corresponds to the Hamiltonian of the Heisenberg spin chain with spin states valued in $\frak{sl}(2)$. On the gauge side, this sector consists out of single trace operators for a single scalar field but including derivatives. In the string side, the resulting strings states are excitations along $AdS_5$-part of the geometry. We refer the reader to \cite{Bellucci:2004qr,Stefanski:2004cw,Park:2005ji} for details. 
The spins are now valued in $\mathfrak{sl}(2)$ and in the coherent state representation 
\begin{align}
|\vec{n}\rangle &= \sum_{m=0}^{\infty} \frac{1}{\cosh\left( \frac{\rho}{2} \right)^{2j}} 
\left( \frac{\Gamma(m + 2j)}{m! \,\Gamma(2j)} \right)^{1/2} 
e^{im\phi} \tanh\left( \frac{\rho}{2} \right)^m |j, j + m \rangle\,, 
\end{align}
they are parametrised by a two-dimensional hyperboloid. In what follows, we fix the spin to live in the infinite dimensional $j=1/2$-representation of $\mathfrak{sl}(2)$. In the continuum limit one again obtains equations of motion which are simply those of the $SU(2)$-sector but adequately analytically continuation \cite{Bellucci:2004qr}
\begin{gather}
\begin{aligned}\label{eq:eqs_cont_sl2}
\sinh \rho \, \partial_t \rho - \frac{\tilde{\lambda}}{L^2} \partial_\sigma \left( \partial_\sigma \phi \, \sinh^2 \rho \right) &= 0, \\
\sinh \rho \, \partial_t \phi + \frac{\tilde{\lambda}}{L^2} \left[ \partial_\sigma^2 \rho - \frac{1}{2} \sinh(2 \rho) \, (\partial_\sigma \phi)^2 \right] &= 0\,,
\end{aligned}
\end{gather}
where $\tilde \lambda=\lambda/(4\pi^2)$.
This corresponds to fast spinning string propagating on a torus $S^1_\phi \times S^1_\varphi$ where the first circle lives in AdS$_5$ and the second in $S^5$.

A simple solution of \eqref{eq:eqs_cont_sl2} corresponding to a rotating string \cite{Bellucci:2004qr} is to take $\partial_\sigma \phi=0$. This implies that $\partial_\tau \rho=0$ and $\partial^2_\tau\phi=0$, implying that $\partial_\tau \phi=\omega$ and $\rho=\rho(\sigma)$ determined by
\begin{align}
    \omega \sinh \rho+\frac{\tilde \lambda}{L^2}\partial^2_\sigma\rho=0\,,
\end{align}
which when integrated once leads to
 \begin{align}
     \partial_\sigma \rho =\pm \sqrt{q-b\cosh \rho}\,,\quad a=\text{cst}\,,\quad b=2L^2\omega/\tilde \lambda\,,
\end{align}
such that $a>b$.
This solution corresponds to a string, with profile $\rho(\sigma)$, rotating at constant angular velocity along an $S^1$ of AdS$_5$.

Correspondingly we obtain a coherent state whose complexity is $\sigma$-dependent
\begin{align}
|\vec{n}\rangle &= \sum_{m=0}^{\infty}  \frac{1}{\cosh\left( \frac{\rho(\sigma)}{2} \right)^{2j}} 
\left( \frac{\Gamma(m + 2j)}{m! \,\Gamma(2j)} \right)^{1/2} 
e^{im\omega \tau} \tanh\left( \frac{\rho(\sigma)}{2} \right)^m|j, j + m \rangle\,, \label{eq:sl_rs}
\end{align}
To determine the complexity associated with the time evolution of the probability distribution \eqref{eq:sl_rs}, we evaluate the autocorrelation function $S(t)$ and calculate the corresponding Lanczos coefficients using the moment recursion method as follows
\begin{equation}
    a_n = -\omega\cosh(\rho(\sigma))\left(n + \frac{L}{2}\right) + \frac{k\omega}{2}, \qquad b_n = \frac{\omega\sinh(\rho(\sigma))}{2}\sqrt{n(n+L-1)}. \label{eq:SL_RS_LC}
\end{equation}
These Lanczos coefficients can be related to those of the $SL(2,\mathbb R)$ coherent state using the parametrisation $\alpha = \omega\sinh(\rho(\sigma))/{2}$, $h = L/{2}$, $\gamma = -\omega\cosh(\rho(\sigma))$, and $\delta = k\omega/{2}$. With this parametrisation, the Lanczos coefficients take the form
\begin{equation}
    a_n = \gamma(h+n) + \delta, \qquad b_n = \alpha\sqrt{n(n+2h-1)}.
\end{equation}
Using the results of \cite{Balasubramanian:2022tpr}, the corresponding spread complexity is then given by
\begin{eqnarray}
    \mathcal{C}(t,\sigma) = \frac{2h}{1-\frac{\gamma^2}{4\alpha^2}} \sinh^2\left(\alpha t \sqrt{1-\frac{\gamma^2}{4\alpha^2}}\right) = \frac{L}{\text{csch}(\rho(\sigma))^2}\sin^2\left(\alpha t \, \text{csch}(\rho(\sigma))\right),
\end{eqnarray}
where $\sqrt{1-\frac{\gamma^2}{4\alpha^2}} = i \, \text{csch}(\rho(\sigma))$ corresponds to the Lanczos coefficients in \eqref{eq:SL_RS_LC}. The resulting complexity oscillates in time, with the oscillation period determined by $\text{csch}(\rho(\sigma))$ and the oscillation amplitude scales with the chain length $L$. 

It is interesting to observe that we obtain an oscillatory complexity in an infinite-dimensional representation. Geometrically, this is understood easily. The string propagates in a compact subspace, namely $S^1\times S^1$ inside a non-compact space, namely $AdS_5\times S^5$.


\subsubsection*{$PSU(1|1)$-subsector}
We now consider a case that includes fermionic degrees of freedom and captures the anomalous dimension of operators of the form  associated to the operators \eqref{eq:operator_su23}. The simplest (all-loop) closed subsector is $PSU(1|1)\subset SU(2|3)\subset PSU(2,2|1)$ \cite{Beisert:2003ys,Beisert:2004ry}, which corresponds to operators with a single scalar and fermionic component $\mathrm{tr}(X^I \psi^K )$, corresponding to a special case of the $SU(2|3)$-single trace operators in  eq. \eqref{eq:operator_su23}.
The equations of motion for this subsector, resulting from the semiclassical approximation and subsequent continuum limit, were obtained and studied in \cite{Hernandez:2004kr} and  \cite{Stefanski:2005tr}. The $PSU(1|1)$-coherent state is given by
\begin{align}
|n\rangle &=  \left( 1 - \frac{1}{2} \zeta^* \zeta \right) |0\rangle + \zeta |1\rangle\,, \label{eq:psu_coh}
\end{align}
where $|0\rangle$ is the bosonic ``vacuum state'' and and $|f\rangle$ is when the state carries the single fermionic state. In a similar way as before, in the continuum limit yields the action
\begin{align}
S &= - \int dt \left[ i \langle n | \frac{\mathrm d}{\mathrm  dt} | n \rangle + \langle n | H | n \rangle \right]= - \frac{L}{2\pi} \int \mathrm  d\sigma \, \mathrm  dt \left[ i \zeta^* \partial_t \zeta + \frac{\lambda}{2L^2} \partial_\sigma \zeta^* \partial_\sigma \zeta \right]\,.\label{eq:psu_ac}
\end{align}
This is the action of a $(1+1)$-dimensional non-relativistic fermion whose equation of motion is
\begin{align}
i \partial_t \zeta = - \frac{L^2}{\lambda}\partial_\sigma^2 \zeta\,.\label{eq:diff_psu}
\end{align}
Since the spin chain is periodic, we impose the boundary condition $\zeta(2\pi, t) = \pm \zeta(0, t),$ leading to the solution 
\begin{align}
\zeta(\sigma, t) = \zeta_0\sum_n \left( A_n e^{-i k_n^2 t} \cos(\kappa_n \sigma) + B_n e^{-i k_n^2 t} \sin(\kappa_n \sigma) \right)\,.
\end{align}
where $\kappa_n = n$ for periodic boundary conditions, $\kappa_n = n + \frac{1}{2}$ for anti-periodic ones and where $k_n^2=L^2\kappa_n^2/\lambda$ and $\zeta_0$ is a unit Grassmann odd vector.  Choosing that at time $t=0$ and, without loss of generality at $\sigma=0$ the coherent state lives in the purely bosonic ground state fixes $A_n=0$ for all $n$ and thus we proceed with
\begin{align}
    \zeta(\sigma, t) = \zeta_0\sum_{n=0}^\infty e^{-i k_n^2 t} \sin(\kappa_n \sigma)\,\label{eq:bc_soln}.
\end{align}
  To find the Krylov basis, first, we notice that the Hamiltonian in the  action \eqref{eq:psu_ac} is Grassmann even and thus cannot change the grading of the state it is acting upon. The coherent state \eqref{eq:psu_coh} for a single lattice site is in a two-dimensional space, and as the initial state is the first Krylov basis vector, the other Krylov basis vector is trivially fixed and is proportional to the only orthogonal vector in that two-dimensional space with the same grading. 
  
  Under the action of the Hamiltonian in the semiclassical action \eqref{eq:psu_ac}, the coherent state \eqref{eq:psu_coh} for a single lattice site thus spans a two-dimensional space. Taking as before the coherent state as the initial state or first Krylov basis vector by determining the only other Krylov basis vector of the same grading.
  In conclusion, the two normalised Krylov basis vectors are
\begin{align}
    \ket{K_0}
    =
    \left( 1 - \frac{1}{2} \bar{\zeta} \zeta \right) |b\rangle + \zeta |f\rangle\,,\qquad \ket{K_1}
    =
    \frac{1}{\gev{\bzeta \zeta}}\left(\bar{\zeta} \zeta |b\rangle - \zeta |f\rangle\right)\,,
\end{align}
where is the semiclassical averaging over fermionic degree of freedom $\gev{-}$, as defined in \eqref{scAverage}. 

Turning back to the general solution in eq. \eqref{eq:bc_soln}, first, we note that the sum converges neither for periodic nor anti-periodic boundary conditions. On the other hand, each individual summand in \eqref{eq:bc_soln} is also a valid solution of \eqref{eq:diff_psu} with appropriate boundary conditions, hence we truncate the series to a finite sum.

The simplest choice of solution is a single summand, namely $\zeta(\sigma, t) = \zeta_0 e^{-i k_n^2 t} \sin(\kappa_n \sigma)$ but this choice renders the term $\bzeta(\sigma, t)\zeta(\sigma, t)$ time-independent and it is straightforward to check that in this case, the coherent state does not span a non-trivial Krylov subspace: the dynamics never spans the Krylov space further than the initial state. 

The simplest, non-trivial choice for $\zeta(\sigma, t)$ has two summands,
\begin{eqnarray}
    \zeta(\sigma, t)=  \zeta_0\sum_{n=1}^2 e^{-i k_n^2 t} \sin(\kappa_n \sigma)\,\label{eq:bc_soln_part}\,.
\end{eqnarray}
This choice results in the following Krylov basis vectors,
\begin{align}
\begin{split}
    | K_0 \rangle
    &=
    \left(1-\frac{\bar{\zeta}_0\zeta_0}{2}\left(\sum_{j=1}^2\sin{(\kappa_{n_j}\sigma)}\right)^2\;\right) \ket{b}
    +
    \zeta_0\left(\sum_{j=1}^2\sin{(\kappa_{n_j}\sigma)}\right)\ket{f}\,, \\ 
    | K_1 \rangle 
    &=
    \bar{\zeta}_0\zeta_0\left(\sum_{j=1}^2\sin{(\kappa_{n_j}\sigma)}\right) \ket{b}-\zeta_0\ket{f}\,.
\end{split}
\end{align}
For the time evolved initial coherent state of \eqref{eq:psu_coh}, we get the spread complexity,
\begin{align}
\begin{split}
    \mathcal{C}(t)
    =
    \gev{|\langle K_1|n(t)\rangle|^2}
    =
    \left(\sum_{j=1}^2\sin{(\kappa_{n_j}\sigma)}\right)
    \left(\sum_{l=1}^2\sin{(\kappa_{n_l}\sigma)}[1-2\cos{(k_{n_l}^2t)}]\right)\,.
\end{split}
\end{align}
This expression can then be generalised to the $L$ lattice sites. However, as was discussed in \ref{subsec:ferm_mf}, the Krylov basis for a general set of displacement parameters is analytically intractable. Instead, one possibility is to specialise to the case where all the different lattice sites have the same parametrisation in the displacement parameter. Indeed, then, as was shown in \ref{subsec:ferm_mf}, the Krylov basis is the uniform linear combination of all permutations of $k$ number of $\ket{K_1}$ states and $L-k$ number of $\ket{K_0}$ states is the $k$th Krylov basis vector. The resulting spread complexity is given by,
\begin{equation}
    \mathcal{C}(t)= |\phi_0^{L-1}\phi_1|^2+2 |\phi_0^{L-2}\phi_1^2|^2+\cdots+N |\phi_1^L|^2=\sum_{k=0}^Lk |\phi_0^{L-k}\phi_1^k|^2\,,
\end{equation}
where we define $\phi_0(t)=\langle K_0|n(t)\rangle$ and $\phi_1(t)=\langle K_1|n(t)\rangle$.

Although, we drew parallels with the calculations in this subsection are similar in nature to those for the single- and multi-fermion cases discussed in section \ref{sec:fermion_osc}, there are several key distinctions as listed in \ref{sec:gen_spread_complex_spinchains}. Nonetheless, we see here the same pattern again: the spread complexity localises into an atomic Krylov path, now dictated by the multi-fermionic HW coherent state discussion in section \ref{subsec:ferm_mf}.

\section{Conclusions and future directions}

In this paper, we extended the concept of spread complexity beyond bosonic systems to include fermionic and supercoherent states. Our central motivation is to generalise the formulation of spread complexity to supersymmetric systems and, in particular, to the AdS/CFT correspondence. Our strategy is to specialise to a sector of holography where we have full control, enabling an analytic extraction of the spread complexity.  Semiclassically, many physical systems are well-approximated in the coherent state path integral formulation. By leveraging the associated algebraic structure, we have computed spread complexity analytically in several systems governed by (super-)groups. 

One of the crucial insights is that the Krylov chain has to be generalised to a \textit{Krylov path} in a higher dimensional lattice from the Lie and super-Lie algebras. This lattice is defined by the weights of the relevant highest weight representation. In particular, the Krylov path explains the appearance of ``emergent dynamical $SL(2)$''-structures, previously observed in \cite{Caputa:2024sux}. We explicitly show this structure in semiclassical systems governed by coherent states.  

Within our framework we compute spread complexity for generic semiclassical Hamiltonian systems. From this point of view, choosing a different Hamiltonian also realises a different trajectory in the phase space of coherent states. Besides identifying the Krylov path, a second crucial ingredient is the trajectory in the phase space of the coherent states. This determines the time dependence of the spread complexity. Note, even though several systems may be governed by the same dynamical $\sl{2}$ or $\su{2}$ spread complexity, their specific Hamiltonians may determine a completely different time dependence and hence complexity. This is mirrored in our results in spread complexity for string states in section \ref{sec:semicl_spin_chains}.

Furthermore, we extended spread complexity to fermionic and multifermionic coherent states~\cite{Fan:1998cb, Adhikari:2023evu}. Our results highlight key differences in their dynamics compared to bosonic systems. Single fermion systems, constrained by their finite-dimensional Krylov subspace, exhibit bounded and periodic complexities, reflecting the limited dimensionality of their Hilbert space. In contrast, multifermion systems present a richer structure due to their extended Hilbert space, resulting in more intricate time evolution shaped by contributions from states with varying total fermion numbers. Constructing the Krylov basis for arbitrary parameterisations of the displacement operator poses significant challenges. To circumvent this, we evaluated the spread complexity for the specific case where all displacement operators share a uniform parameterisation and find an oscillatory complexity. For the general case, we derive an upper bound for the spread complexity, calculated in the Fock basis of the multifermion system. We achieve this by taking randomly chosen coefficients for displacement parameters. The resulting collective behaviour of this multifermion system demonstrates a non-oscillatory upper bound of the spread complexity, driven by the combined effects of individual mode contributions. 

We then consider superalgebras and their associated coherent states. We provided explicit realisations of Krylov paths through a lattice, whose directions are quantified by bosonic and fermionic degrees of freedom. Evolution through these lattices are generated by displacement operators which combine bosonic and fermionic ladder operators. We argued that spread can be quantified in two ways:
 \begin{itemize}
     \item[1)] The conventional spread complexity $\cC$ accounts for the spread through the Krylov basis. The Krylov basis and Krylov operator emerge from a \textit{dynamical algebra} characterising the displacement operator which drives the evolution.
     \item[2)] Viewing the Krylov chain as a path through the lattice, we can use the lattice basis vectors to quantify spread with respect to the lattice itself.
 \end{itemize}
The examples we considered are the super HW algebra and $\osp{2|1}$ superalgebra. When choosing the Grassmann parity of the exponents in the displacement operators to be even, we can apply general considerations for coherent states following from supergroup analyses. We furthermore looked at evolution driven by a Grassmann odd Hamiltonian given by the supercharge. Interestingly, we were able to relate its spread to complexities derived from Laguerre polynomials, which had been derived on mathematical grounds in \cite{muck2022krylov}. Our analysis provides a link to a physical realisation thereof in supersymmetric systems.

In the last part, building on these insights, we considered superstring states and their dual operators in the semiclassical regime of planar holography. In this regime of the correspondence, the dynamics of the string states and dual operators are \textit{equivalent}. In particular, the computed complexities are valid on both sides of the duality. Working with the continuum limit of the spin chain implies in particular that spread complexity depends not only on time but also on space. Taking as initial state a coherent state on every site of the spin chain and driving their evolution with the spin chain Hamiltonian, for different subsectors of $PSU(2,2|4)$, we obtain the corresponding spread complexities. Holographically, this corresponds to superstring states propagating on submanifolds of $AdS_5\times S^5$.

Despite being far remote from the examples considered in earlier section, the spread complexity assumes once more the atomic form of an $\sl{2}$ or $\su{2}$ spread complexities. 
Crucially however, the spin chain Hamiltonian realises different trajectories in the corresponding coherent state phase spaces, thus leading to spread complexities with different time and space dependences.

The results we obtain suggest that spread complexity captures the geometry in which the string propagates. In particular, bounded complexities reflect compactness of the target space submanifold explored by the string, even when embedded in the non-compact part of $AdS_5\times S^5$.

\subsubsection*{Future directions}

In this article, we have only scratched the surface of a wide range of applications and exploration of spread complexity to semiclassical systems harbouring fermions and bosons. Below we list a number of them.

\begin{itemize}
    \item{\bf Beyond rotating strings and one-loop.} Although we have limited the discussion to rotating strings, it is well known that the coherent state -- propagating string duality admits a wide variety of solutions, including e.g. pulsating, spiky strings or magnons. In addition, these solutions often survive integrable or non-integrable deformation, either induced by including higher-derivative interactions \cite{Kruczenski:2004kw,McLoughlin:2022jyt} or deformation of the target space \cite{Roiban:2003dw,Zoubos:2010kh}. It would be interesting to compute spread complexity for these example and assess whether spread complexity could be used to differentiate between chaos and integrability in this semiclassical limit of holography. It would be interesting to contrast this with integrable-chaotic transitions \cite{Rabinovici:2022beu} witnessed by Krylov complexity for operators in the spin chain.   
    \item{\bf Topology and Lin-Lunin-Maldacena geometries.} Another sector of the holographic correspondence, which is amenable in the semiclassical limit, is given by Lin-Lunin-Maldacena geometries \cite{Lin:2004nb}. Indeed, in the half-BPS sector, these smooth geometries are semiclassically dual to coherent state in the dual CFT. Lin-Lunin-Maldacena geometries form a rich family of supergravity solutions featuring different topological structures \cite{Berenstein:2005aa}. Using the framework above, a coherent state of a graviton can be used to probe the topology of the gravity background. We plan to report on this soon.
    \item{\bf Fractional statistics and anyonic chains.} The physical world is not limited to states of even or odd statistics but is known to include states of fractional statistics. Here again, our approach offers a clear path to elucidate the generalise spread complexity to systems with fractional symmetry and anyonic states. Notable examples are the anyonic Hubbard model \cite{keilmann2011statistically} or the effective description of double-scaled SYK \cite{Berkooz:2018jqr}, which could present a rich framework for investigating the effects of anynonic commutation relations on complexity measures. 
    \item{\bf Higher rank algebras and Krylov path.} A crucial new element in many of our examples was to consider Lie algebras of rank larger than one. In those examples, we elucidated how spread complexity systematically reduces to an effective rank-one Lie algebra. The latter has a simple physical interpretation: it generates the Krylov path following the Hamiltonian time evolution of the initial state through the Hilbert space parametrised  by the weight lattice. These insights deserve a systematic study, enabling us to tackle spread complexity in systems such as, e.g., higher-dimensional CFTs which could then be compared to the results of \cite{Chagnet:2021uvi}. More fundamentally, this may enlighten the particular role of the emergent $SL(2,\mathbb R)$ and $SU(2)$ guiding spread complexity trajectories remains an area of fundamental interest. Understanding the universality of these structures across different systems could significantly.
 \end{itemize}

\subsection*{Acknowledgements}
We thank Shira Chapman for her contribution in early stages of the project. We thank Aranya Bhattacharya and Axel Kleinschmidt for useful discussions. All of us are supported by the German Research Foundation (DFG) through a German-Israeli Project Cooperation (DIP) grant ``Holography and the Swampland''.
RND~and JE~are supported by Germany's Excellence Strategy through the W\"urzburg‐Dresden Cluster of Excellence on Complexity and Topology in Quantum Matter ‐ ct.qmat (EXC 2147, project‐id 390858490).
RND~further acknowledges the support by  Deutscher Akademischer Austauschdienst (DAAD, German Academic Exchange Service) through the funding programme \enquote{Research Grants - Doctoral Programmes in Germany, 2021/22 (57552340)}. 
CN's work is funded by the European Union’s Horizon Europe Research and Innovation Programme under the Marie Skłodowska-Curie Actions COFUND, Physics for Future, grant agreement No 101081515.
SD is supported by an Azrieli fellowship funded by the Azrieli foundation. CN and SD are supported by the Israel Science Foundation
(grant No. 1417/21),  by Carole and Marcus Weinstein through the BGU Presidential Faculty Recruitment Fund, by the ISF Center of Excellence for theoretical high energy physics, and by the ERC Starting Grant dSHologQI (project number 101117338).

\newpage
\appendix

\section{Lanczos algorithm and return amplitude}\label{app:Lanczos_survival}
In this appendix we succinctly review to methods to obtain the Lanczos coefficients, via the Lanczos algorithm and the return amplitude. We provide in addition details to the applying the Lanczos algorithm to fermionic Hamiltonian in subsection \ref{OSPGrassmannOdd}. 
\subsection{The conventional Lanczos algorithm}
Using the natural inner product on $\cH$, we can orthonormalize \eqref{eq:basis state} through the Lanczos algorithm: 
\begin{itemize}
\item[1.] $b_0\equiv0$,
\item[2.] $|K_0\rangle \equiv |\psi(0)\rangle, a_0=\langle K_0|H|K_0\rangle$
\item[3.] For $n\geq1$: $|\mathcal{A}_n\rangle=(H-a_{n-1})|K_{n-1}\rangle-b_{n-1}|K_{n-2}\rangle$
\item[4.] Set $b_n=\sqrt{\langle \mathcal{A}_n|\mathcal{A}_n\rangle}$
\item[5.] If $b_n=0$ stop; otherwise set $|K_n\rangle=\frac{1}{b_n}|\mathcal{A}_n\rangle,~a_n=\langle K_n|H|K_n\rangle$\,, and repeat step 3.
\end{itemize}
If $\mathcal{D}$ is finite, then this Lanczos algorithm ends with $b_\mathcal{D}=0$. The resulting orthonormal basis ${|K_n\rangle}_{n=0}^{\mathcal{D}-1}$ is called the Krylov basis.

\subsection{The Lanczos coefficients from the return amplitude}
An alternative way to calculate spread complexity is by the moment recurrence method. For a given return amplitude, $S(t) = \langle \psi(t) | \psi(0) \rangle$, the $n$-th moment of the return amplitude can be derived by taking the $n$-th derivative of $S(t)$ and evaluating it in the limit $t \to 0$. This procedure is formalised as follows~\cite{Parker:2018yvk, Balasubramanian:2022tpr}
\begin{align}
    \mu_n = \frac{1}{i^n} \lim_{t \to 0} \frac{d^n }{dt^n}S(t) \,. \label{eq:moments}
\end{align}
The zeroth moment, $\mu_0 = S(0)$, corresponds to the value of the return amplitude at the initial time. By construction, the return amplitude is normalised such that $S(0) = 1$, leading to $\mu_0 = 1$. 
Using these moments, the Lanczos coefficients, $a_n$ and $b_n$, can be computed through a recursive algorithm~\cite{viswanath1994recursion}. The algorithm involves defining intermediate quantities $\mathsf{M}_k^{(n)}$ and $\mathsf{L}_k^{(n)}$, and proceeds as follows
\begin{align}
    \mathsf{M}_k^{(0)} &= (-1)^k \mu_k \,, \quad \mathsf{L}_k^{(0)} = (-1)^{k+1} \mu_{k+1} \,, \nonumber \\
    \mathsf{M}_k^{(n)} &= \mathsf{L}_k^{(n-1)} - \mathsf{L}_{n-1}^{(n-1)} \frac{\mathsf{M}_k^{(n-1)}}{\mathsf{M}_{n-1}^{(n-1)}} \,, \nonumber \\
    \mathsf{L}_k^{(n)} &= \frac{\mathsf{M}_{k+1}^{(n)}}{\mathsf{M}_n^{(n)}} - \frac{\mathsf{M}_k^{(n-1)}}{\mathsf{M}_{n-1}^{(n-1)}} \,, \quad k \geq n \,. \nonumber
\end{align}
Finally, the Lanczos coefficients are determined as
\begin{align}
    b_n &= \sqrt{\mathsf{M}_n^{(n)}} \,, \quad a_n = -\mathsf{L}_n^{(n)} \,. \label{eq:lanczos}
\end{align}
After obtaining the Lanczos coefficients, one can solve Schrödinger's equation in the Krylov space \eqref{eq:Krylov chain for state} and obtain the Krylov space wave functions and the corresponding complexity.

\subsection{Lanczos algorithm for supercharge evolution}
\label{appSusyKrylov}
In this appendix, we perform Lanczos algorithm for the fermionic operator in eq. \eqref{SusyLiouvillian}.

Let $\ket{K_0}=\ket{h}$ be the state to be evolved. We employ the following notation for the norm of a vector in Hilbert space $\norm{\ket{v}}=\sqrt{\braket{v}}$. Keep in mind that $G_{\pm1/2}^\dagger=G_{\mp1/2}$. The first iteration of the Lanczos algorithm gives 
\begin{align}
    \ket{A_1}
    &=
    \cL\ket{h}
    =
    \alpha G_{-1/2}\ket{h}\,,
    \\
    b_1^2
    &=
    \alpha^2\norm{G_{-1/2}\ket{h}}^2
    =
    2\alpha^2h
    \\
    \ket{K_1}
    &=
    \frac{1}{b_1}G_{-1/2}\ket{h}
    =
    \frac{1}{\norm{G_{-1/2}\ket{h}}}G_{-1/2}\ket{h}
\end{align}
The second iteration gives
\begin{align}
    \ket{A_2}
    &=
    \frac{\alpha}{\norm{G_{-1/2}\ket{h}}} G_{-1/2}^2\ket{h}
    \\
    b_2^2
    &=
    \alpha^2\frac{\norm{G_{-1/2}^2\ket{h}}^2}{\norm{G_{-1/2}\ket{h}}^2}
    =
    2\alpha^2h
    \\
    \ket{K_2}
    &=
    \frac{1}{b_2}G_{-1/2}^2\ket{h}
    =
    \frac{1}{\norm{L_{-1}\ket{h}}}L_{-1}\ket{h}
\end{align}
where $G_{-1/2}^2=L_{-1}$ has been used. By running more iterations the following pattern emerges for the Krylov basis
\begin{equation}\label{SusyKrylov}
    \ket{K_n}=\frac{G_{-1/2}^n\ket{h}}{\norm{G_{-1/2}^n\ket{h}}},
    \qquad
    b_n=\alpha\frac{\norm{G_{-1/2}^n\ket{h}}}{\norm{G_{-1/2}^{n-1}\ket{h}}}
\end{equation}
Evidently these states are all orthogonal since they lie in distinct energy eigenspaces, each shifted from their neighbor in half-integral steps. The Lanczos coefficients $b_n$ trade the normalisation of the preceding Krylov basis element $\ket{K_{n-1}}$ for that of $\ket{K_n}$, as they should.

In order to evaluate the Lanczos coefficients in \eqref{SusyKrylov} in generality, we require the norms $\norm{G_{-1/2}^n\ket{h}}^2=\bra{h}G_{1/2}^nG_{-1/2}^n\ket{h}$. For even integers $n=2k$ this is easily achieved by virtue of $G_{-1/2}^2=L_{-1}$,
\begin{equation}
    \norm{G_{-1/2}^n\ket{h}}^2
    =
    \norm{G_{-1/2}^{2k}\ket{h}}^2
    =
    \norm{L_{-1}^k\ket{h}}^2
    =
    k!\frac{\Gamma(2h+k)}{\Gamma(2h)}\,.
\end{equation}
A little sidestep is necessary for odd $n=2k+1$. The state
\begin{equation}
    G_{-1/2}\ket{h}=\eta\ket{h+1/2}
\end{equation}
has some normalisation $\eta\in \mathbb{C}$ that we need to compute. The state $\ket{h+1/2}$ is taken to have unit norm, just as $\ket{h}$. Then
\begin{equation}
    \norm{G_{-1/2}\ket{h}}^2
    =
    \bra{h}\{G_{1/2},G_{-1/2}\}\ket{h}
    =
    2h
    \overset{!}{=}|\eta|^2\,,
\end{equation}
hence we can take $\eta=\sqrt{2h}\in\mathbb{R}$,
\begin{equation}
    G_{-1/2}\ket{h}=\sqrt{2h}\ket{h+1/2}\,.
\end{equation}
Note that this state is still an $\textrm{SL}(2,\mathbb{R})$ lowest weight state, $L_{1}G_{-1/2}\ket{h}=0$. The norm for odd $n=2k+1$ is thus
\begin{align}
    \norm{G_{-1/2}^{n}\ket{h}}^2
   =
    2h\norm{L_{-1}^{k}\ket{h+1/2}}^2
    =
    k!\frac{\Gamma(2h+k+1)}{\Gamma(2h)}
\end{align}
In conclusion, the Lanczos coefficients are for even $n=2k$ and odd $n=2k+1$, respectively
\begin{equation}\label{OSP12Lanczoscoeff}
    b_{2k}
    =
    \alpha\frac{\norm{G_{-1/2}^{2k}\ket{h}}}{\norm{G_{-1/2}^{2k-1}\ket{h}}}
    =
    \alpha\sqrt{k}\,,
    \qquad
    b_{2k+1}
    =
    \alpha^2\frac{\norm{G_{-1/2}^{2k+1}\ket{h}}}{\norm{G_{-1/2}^{2k}\ket{h}}}
    =
    \alpha\sqrt{2h+k}\,
\end{equation}
as discussed in the main text.

\section{Additional material on $\osp{2|1}$}
In this appendix, we provide additional details on the material presented in section \ref{sec:OSP}.

\subsection{On the presence of the $R$-charge $J$}\label{appJ}

The presence of the \quotes{fermion number} operator $J$ for our discussion in subsection \ref{sec:OSP} can be made precise by considering the larger supergroup $\osp{2|2}\supset\osp{2|1}$ generated by the eight elements $(L_0, L_{\pm1}, J, G^{a}_r)$ where $a=\pm$ and $r=\pm1/2$. Its Cartan subalgebra consists of the two bosonic operators $(L_0,J)$ and is thus of rank two. The former operator is an $\mathfrak{sl}(2)$ charge as before, and the latter is responsible for a $U(1)$ R-Symmetry. The fermionic charges $G^{a}_r$ string $\mathfrak{sl}(2)$ representations together.

These $\osp{2|2}$ representations are in general larger than the one studied in \eqref{GradedKrylovSpace}. Ground states of highest weight representations are annihilated by $L_1$ and $G^\pm_{1/2}$. There is a special type of representation however, called `atypical' or `(anti-)chiral', whose ground state is annihilated additionally by $G_{-1/2}^+\,(G_{-1/2}^-)$. Hence only one supercharge operator, namely $G_{-1/2}^-\,, (G_{-1/2}^+)$ acts non-trivially on this ground state. It follows that chiral representations look precisely like \eqref{GradedKrylovSpace}. Furthermore, only the operators $L_0,L_{\pm1},G^-_{\pm1/2}$ and $J$ act non-trivially on this representation. By identifying $G^-_{\pm1/2}\to G_{\pm1/2}$ we return to the situation studied in this section, with the additional $R$ charge $J$ in hand. It has the desired properties of disregarding the $\mathfrak{sl}(2)$ subalgebra and counting the supercharge with one (negative\footnote{Had we picked an anti-chiral representation, the R-charge would count the supercharge with a positive unit.}) unit
\begin{equation}
    [L_0,J]=0,
    \quad
    [L_{\pm1},J]=0,
    \quad
    [J,G_{\pm1/2}]=-G_{\pm1/2}
\end{equation}
It can thus only detect spread in the vertical direction of figure \ref{fig:SUSY_hwrep} and acts on the ground state of \eqref{GradedKrylovSpace} by $J\ket{h}=2h\ket{h}$.

\subsection{$\osp{2|1}$ spread through weight lattice spanned by $(L_0,J)$}\label{app:spreadThroughWeightLattice}
In order to study spread induced by an $\osp{2|1}$ displacement operator in the weight lattice of \eqref{GradedKrylovSpace}, have to express the coherent state \eqref{ospDisplacer} state in the eigenbasis of $(L_0,J)$ found in \eqref{sl2Kbasis}. This is done by factorising the state as follows \cite{fatyga1989baker} 
\begin{align}
D(\xi,\chi)&=
\exp\left[\xi L_{-1}-\bxi L_{1}+\chi G_{-\frac{1}{2}}+\bchi G_{\frac{1}{2}}\right]=
e^{\beta L_{-1}}e^{\delta G_{-1/2}}e^{\gamma L_{1}}e^{\epsilon G_{1/2}}e^{\alpha L_{0}}\label{factorizedD}
\end{align}
The parameters on the right-hand side are expanded in terms of the Grassmann numbers $\chi$
\begin{subequations}\label{GradechiCoords}
\begin{align}
e^{\alpha}
&=
e^{\alpha_0}(1-\alpha_1\,\bchi\chi)
=
    \frac{1}{\cosh^2(|\xi|)}\left(1-2\frac{|\xi|\tanh(|\xi|)+\cosh^{-1}(|\xi|)-1}{|\xi|^2}\bchi\chi\right)\\
\beta
&=
\beta_0-\beta_1\,\bchi\chi =
  \frac{\xi}{|\xi|}\tanh(|\xi|)
  +\frac{\xi}{|\xi|^3}\frac{|\xi|-\sinh(|\xi|)}{\cosh^2(|\xi|)}\bchi\chi\\
\delta
&=
\delta_0\chi+\delta_1\bchi =
    \frac{\tanh(|\xi|)}{|\xi|}\chi+\frac{\xi}{|\xi|^2}\left(1-\frac{1}{\cosh(|\xi|)}\right)\bchi
\end{align}
\end{subequations}
and the functions $\epsilon(\xi,\chi), \gamma(\xi,\chi)$ are not needed, but can nevertheless be found in \cite{fatyga1989baker}. For $\chi=0$, the $SL(2,\R)$ displacement operator in \cite{Caputa:2021sib} is recovered. Observe that the displacement operator in the supersymmetric Heisenberg-Weyl algebra of subsection \ref{sec:coh_superHW} factorises neatly into bosonic and fermionic components. This is in contrast to the current example, where the bosonic and fermionic parameters mix when factorising the displacement operator as evident from \eqref{GradechiCoords}. This is the source of interesting spread through Hilbert space in this case.

We now evolve the $OSp(2|1)$ lowest weight vector $\ket{h}$ by the displacement operator \eqref{factorizedD},
\begin{align}
    \ket{\xi,\chi,h}
    :=
    D(\xi,\chi)\ket{h}
    &=
    e^{\alpha h}e^{\beta L_{-1}}e^{\delta G_{-1/2}}\ket{h}\notag\\
    &=
    \sum_{n=0}^\infty\left(\phi^h_{n}\ket{h,n}+\phi_n^{h+1/2}\ket{h+1/2,n}\right)
\end{align}
where we have used $G_{-1/2}\ket{h}=\sqrt{2h}\ket{h+1/2}$, $L_{-1}^n\ket{h}=\sqrt{\frac{n!\Gamma(2h+n)}{\Gamma(2h)}}\ket{h,n}$, with normalised states $\braket{h,n}{h',m}=\delta_{h+n,h'+m}$ and the nilpotency of the Grassmann odd elements $\chi^2=\bchi^2=0$. It is important to stress that the basis $\ket{h,n}$ is \textit{not} the Krylov basis, and thus the wave functions
\begin{align}
    \phi_n^{(h)}
    =
    e^{\alpha h}\beta^n\sqrt{\frac{\Gamma(2h+n)}{n!\,\Gamma(2h)}}\,,
    %
    %
    \qquad
    \phi_n^{(h+1/2)}
    =
    e^{\alpha h}\beta^n\,\delta\,\sqrt{\frac{\Gamma(2h+1+n)}{n!\,\Gamma(2h)}}
\end{align}
are neither the Krylov wave functions. They still encode probabilities however. To see this we first evaluate their absolute values
\begin{subequations}\label{OSPpreProb}
\begin{align}
    |\phi_n^{(h)}|^2
    &=
    e^{2\alpha_0 h}|\beta_0|^{2n}\left[1-\left(2\alpha_1h+n\frac{\beta_1^*\beta_0+\beta_0^*\beta_1}{|\beta_0|^2}\right)\bchi\chi\right]\frac{\Gamma(2h+n)}{n!\,\Gamma(2h)}\\
    |\phi_n^{(h+1/2)}|^2
    &=
    e^{2\alpha_0 h}|\beta_0|^{2n}\left(|\delta_0|^2-|\delta_1|^2\right)\,\bchi\chi\,\frac{\Gamma(2h+1+n)}{n!\,\Gamma(2h)}
\end{align}
\end{subequations}
As discussed in section \ref{secSuperDisplacers}, due to the presence of the Grassmann odd contributions, these are only `pre-probabilities'. Note however that their series
\begin{align}\label{preProbSum}
    \sum_{n=0}^\infty|\phi_n^{(h)}|^2
    &=
    1-4h\frac{\cosh(|\xi|)-1}{|\xi|^2}\bchi\chi\,,\qquad 
    \sum_{n=0}^\infty|\phi_n^{(h+1/2)}|^2
    =
    4h\frac{\cosh(|\xi|)-1}{|\xi|^2}\bchi\chi
\end{align}
already sum up to one. To construct proper probabilities, we must average over the Grassmann odd parameters as in \eqref{scAverage}. Eventually we are interested in scenarios where $\chi=f(t)\zeta_0$ for some function $f(t)\in \mathds{C}$, and $\zeta$ is a Grassmann-valued unit vector, so that 
\begin{align}
    p_n^{(h)}
    &=
    \int d\zeta_0 d\bar\zeta_0 e^{\bar\zeta_0\zeta_0}|\phi_n^{(h)}|^2\,,\qquad 
    p_n^{(h+1/2)}
    =
    \int d\zeta_0 d\bar\zeta_0 e^{\bar\zeta_0\zeta_0}|\phi_n^{(h+1/2)}|^2
\end{align}
which yield the same expressions as in \eqref{OSPpreProb} with $\bchi\chi$ replaced by $|f(t)|^2$. It is convenient to define the probabilities of being in the sector $\cH^{\mathfrak{sl}(2)}_h$ or $\cH^{\mathfrak{sl}(2)}_{h+1/2}$, respectively,
\begin{equation}\label{ProbSectors}
    P^{(h)}
    =
   \sum_{n=0}^\infty p_n^{(h)}\,,\qquad 
    P^{(h+1/2)}
    =
   \sum_{n=0}^\infty p_n^{(h+1/2)}
\end{equation}
which yield the same expressions as in \eqref{preProbSum} with $\bchi\chi$ replaced by $|f(t)|^2$.

As discussed in subsection \ref{secSuperDisplacers}, in order to compute the complexity we require the symbol
\begin{align}
    \bra{\xi,\chi,h}L_0-h\ket{\xi,\chi,h}\label{eq:OSP12preComplexity}
    &=
    \sum_{n=0}^\infty \left(n|\phi_n^{(h)}|^2
    +
    \left(\frac{1}{2}+n\right)|\phi_n^{(h+1/2)}|^2
    \right)\notag\\
    &=
    2h\sinh^2(|\xi|)
    +
    \frac{2h}{|\xi|^2}\bigl(\cosh(|\xi|)-\cosh(2|\xi|)+|\xi|\sinh(2|\xi|)\bigr)\bchi\chi
\end{align}
The  spread complexity is then extracted upon semiclassical averaging using \eqref{scAverage}, which we do upon choosing $\chi=f(t)\zeta_0$, leading to the complexity
\begin{gather}
\begin{aligned}
   \cC_{\textrm{osp}(1|2)}^{(h)}(\xi,\chi) 
   &=\int d\zeta_0 d\bzeta_0\, e^{\bzeta_0\zeta_0}\,\bra{\xi,\chi,h}L_0-h\ket{\xi,\chi,h}
   \label{eq:OSP12cplxApp}\\
   &=
   2h\sinh^2(|\xi|)
    +
    \frac{2h}{|\xi|^2}\bigl(\cosh(|\xi|)-\cosh(2|\xi|)+|\xi|\sinh(2|\xi|)\bigr)|f(t)|^2\,.
\end{aligned}    
\end{gather}
Once more, had we only considered an $SL(2,\R)$ evolution, i.e. with $\chi=0$, the second term would have been missing and we had recovered the expected result $\cC_{\textrm{SL}(2,\mathbb{R})}^{(h)}(\xi)=2h\sinh^2(|\xi|)$. This result is reported in equation \eqref{eq:OSP12cplx} of the main text.

 The symbol of $J$ is 
\begin{align}
    \bra{\xi,\chi,h}J-2h\ket{\xi,\chi,h}\label{eq:OSP12preComplexityJApp}
    &=
    -\sum_{n=0}^\infty 
    |\phi_n^{(h+1/2)}|^2
\end{align}
Using $\chi=f(t)\chi_0$, the prescription \eqref{SpreadCplxSusy} provides the complexity quantifying spread in the vertical direction of the lattice in figure \ref{fig:SUSY_hwrep},
\begin{align}\label{JspreadApp}
    \cC_{\osp{2|1}}^{J}
    =
    P^{(h+1/2)}
    =
    4h\frac{\cosh(|\xi|)-1}{|\xi|^2}|f(t)|^2
\end{align}
We find that it is measured by the probability of residing in the sector $\cH^{\mathfrak{sl}(2)}_{h+1/2}$, which is defined in \eqref{ProbSectors}. This result is reported in equation \eqref{Jspread} of the main text.

\section{Fermionic equations of motion}\label{app:FermionSymplectic}
In this appendix we construct a semiclassical theory for fermionic theories, whose equations of motion indicate how Grassmannian parameters evolve.

\subsection{Single fermion displacement operator with Grassmann odd parameter}
Given the coherent states \eqref{fermionCoherentState} we can generate a semiclassical theory in the following way. First we construct the symbols
\begin{align}
    \bra{\theta}a_F\ket{\theta}=\theta,
    \qquad
    \bra{\theta}a_F^\dagger\ket{\theta}=\bar\theta
\end{align}
Using these and a Grassmannian variable $\eta$, a Hamilton function can be constructed,
\begin{equation}
    H
    =
    \bra{\theta}a_F^\dagger\eta+\bar\eta a_F\ket{\theta}
    =
    \bar\theta\eta+\bar\eta\theta
\end{equation}
A symplectic form is furthermore generated from the return amplitude $|\varphi_0|^2=|\braket{0}{\theta}|^2=(1-\bar\theta\theta)$ as follows. First,  Grassmannian differentials are introduced
\begin{equation}
    \delta=d\theta\,\partial_\theta
    \qquad
    \bar\delta
    =d\bar\theta\,\partial_{\bar\theta}
\end{equation}
The symplectic form is then
\begin{align}
    \omega
    =
    i\delta\wedge\delta \log |\varphi_0|^2
    =
    -i\delta\wedge d\bar\theta\, \theta
    =
    +id\bar\theta\wedge d\theta(\partial_\theta \theta)
    =
    id\bar\theta\wedge d\theta
\end{align}
where $\log(1+\bar\theta\theta)=\bar\theta\theta$ has been used. In contrast to bosonic symplectic forms, fermionic ones are symmetric as opposed to antisymmetric, i.e. $d\bar\theta\wedge d\theta=d\theta\wedge d\bar\theta$ since interchanging the differentials picks up one sign from the wedge product, but also one from interchanging Grassmannian variables. Thus as a matrix in $d\bar\theta$ and $d\theta$ the symplectic form is represented by a Pauli matrix, $\omega=i\sigma_x$, so that its inverse is $\omega^{-1}=-i\sigma_x$. 

Poisson brackets are given by $\{F,G\}=\omega^{\mu\nu}(\partial_\mu F)(\partial_\nu G)$. The equation of motion is thus 
\begin{align}
    \dot{\theta}
    =
    \{\theta,H\}
    =
    \omega^{\theta\bar\theta}(\partial_{\bar\theta}H)
    =
    \eta
\end{align}
and a similar equation for the complex conjugate $\bar\theta$. 

The solution is clearly $\theta=i\eta t+\chi$, where $\chi$ is a Grassmannian constant. For evolution with the displacement operator $D(\theta(t))=\exp(a_F\theta-\bar\theta a_F^\dagger)$, we require that $D(\theta(0))=\mathbf{1}$ so that $\ket{\theta=0}=\ket{0}$. This fixes $\chi=0$. Furthermore, $\eta$ is proportional to a unit Grassmannian vector $\zeta_0$, i.e. $\eta=\alpha\zeta_0$. We arrive at $\theta=i\alpha\zeta t$ as claimed in the main text.

\newpage

\bibliographystyle{JHEP}
\bibliography{refs}

\end{document}